\documentclass[10pt,journal,a4paper,twoside,twocolumn]{IEEEtran}

\IEEEoverridecommandlockouts
\usepackage{cite}
\usepackage{float}
\usepackage{stfloats}
\usepackage{fancyhdr}
\usepackage{setspace}
\usepackage{color}
\usepackage{amsmath,amsfonts,amssymb}
\usepackage{graphicx}
\graphicspath{{./figures/}}%˵�������ص�ͼƬ��·�����ں���ͼƬ���ص�ʱ������д·����ֱ����ͼƬ����
\usepackage{algorithm}
\usepackage{algorithmic}
\usepackage{epsfig}
\usepackage{subfigure}
\usepackage{epstopdf}
\usepackage{txfonts}
\usepackage{bm}
\usepackage{array}
\usepackage{bbm}

\newtheorem{lemma}{Lemma}
\newtheorem{Rem}{Remark}

\begin{document}
%	\pagenumbering{number}
	\title{
	 Massive Wireless Energy Transfer without Channel State Information via Imperfect Intelligent Reflecting Surfaces
	}
	\author{\IEEEauthorblockN{
			Cheng Luo, \emph{Student Member, IEEE}, Jie Hu, \emph{Senior Member, IEEE}, Luping Xiang, \emph{Member, IEEE}, Kun Yang, \emph{Fellow, IEEE}, and Kai-Kit Wong \emph{Fellow, IEEE}
            }
			\\
		% \thanks{xx are with the School of Information and Communication Engineering, University of Electronic Science and Technology of China, Chengdu 611731, China, email: 

		% }
        \thanks{Cheng Luo, Jie Hu, and Luping Xiang are with the School of Information and Communication Engineering, University of Electronic Science and Technology of China, Chengdu, 611731, China, email: chengluo@std.uestc.edu.cn; hujie@uestc.edu.cn; luping.xiang@uestc.edu.cn.}
        % \thanks{Kun Yang is with the Yangtze Delta Region Institute (Quzhou), University of Electronic Science and Technology of China, Quzhou 324000, China, and also with the School of Information and Communication Engineering, University of Electronic Science and Technology of China, Chengdu 611731, China, email: kunyang@uestc.edu.cn.}
        \thanks{Kun Yang is with the School of Computer Science and Electronic Engineering, University of Essex, Colchester CO4 3SQ, U.K., email: kunyang@essex.ac.uk}
        \thanks{Kai-Kit Wong is with the Department of Electronic and Electrical Engineering, University College London, WC1E 6BT, London, U.K., email: kai-kit.wong@ucl.ac.uk.}
	}
	\maketitle
	\thispagestyle{fancy} 
	\lhead{} 
	\chead{} 
	\rhead{} 
	\lfoot{} 
	\cfoot{} 
	\rfoot{\thepage} 
	\renewcommand{\headrulewidth}{0pt}
	\renewcommand{\footrulewidth}{0pt} 
	\pagestyle{fancy}
    \rfoot{\thepage}
	\begin{abstract}
		Intelligent Reflecting Surface (IRS) utilizes low-cost, passive reflecting elements to enhance the passive beam gain, improve Wireless Energy Transfer (WET) efficiency, and enable its deployment for numerous Internet of Things (IoT) devices. However, the increasing number of IRS elements presents considerable channel estimation challenges. This is due to the lack of active Radio Frequency (RF) chains in an IRS, while pilot overhead becomes intolerable. To address this issue, we propose a Channel State Information (CSI)-free scheme that maximizes received energy in a specific direction and covers the entire space through phased beam rotation. Furthermore, we take into account the impact of an imperfect IRS and meticulously design the active precoder and IRS reflecting phase shift to mitigate its effects. Our proposed technique does not alter the existing IRS hardware architecture, allowing for easy implementation in the current system, and enabling access or removal of any Energy Receivers (ERs) without additional cost. Numerical results illustrate the efficacy of our CSI-free scheme in facilitating large-scale IRS without compromising performance due to excessive pilot overhead. Furthermore, our scheme outperforms the CSI-based counterpart in scenarios involving large-scale ERs, making it a promising solution in the era of IoT.
	\end{abstract}
\section{Introduction}
    The successful deployment of Internet of Things (IoT) devices in sixth-generation (6G) wireless communication systems relies heavily on the advancement of Wireless Energy Transfer (WET) technologies. The use of a large number of IoT devices in various high-maintenance environments such as healthcare, environmental detection, smart homes, smart cities, autonomous vehicles, and national defense requires efficient and reliable WET solutions \cite{intro_overviewbruno1,intro_overviewbruno2,SWIPT3,intro_Mon,huj6G}. In addition, the future density of IoT devices is projected to increase to tens or more per square meter \cite{intro_overviewonel1}. As frequent maintenance of batteries in IoT devices is not practical, the development of WET technology is crucial for the successful implementation of IoT in 6G wireless communication systems. Several studies have highlighted the significance of WET in IoT \cite{mahmood2020white,lopez2020csi, zhangBX1}, emphasizing the need for continued research in this field.

    Although the concept of WET is not new, it has been in development for over a century, with one of the earliest experiments conducted by Tesla in 1891. Since then, there have been numerous advancements in WET and Wireless Information Transfer (WIT) technologies, including Simultaneous Wireless Information and Power Transfer (SWIPT), Wirelessly Powered Communication Networks (WPCN), and Wirelessly Powered Backscatter Communication (WPBC), etc \cite{SWIPT1,YueSWIPT,WPCN1,WPBC1}. Despite these developments, limitations in terms of transmission distance and efficiency still exist. Ongoing research is focused on addressing these limitations and developing more efficient and reliable WET solutions.

    Intelligent Reflecting Surface (IRS) is an emerging technology for 6G wireless networks that has the potential to enhance energy efficiency, coverage, and security, while also presenting new opportunities and challenges for SWIPT/WPCN/WPBC\cite{IRSassistedSWIPT,IRSassistedSWIPT2,IRSassistedSWIPT5,IRSassistedWPCN}. An IRS consists of a planar surface with a large number of low-cost, passive reflecting elements, each of which can independently adjust the phase and amplitude of incident electromagnetic signals to meet specific functional and performance requirements. IRS can operate in Full Duplex (FD) mode without self-interference, and can reflect signals to users who are blocked from communicating directly with the Power Beacon (PB). This is a promising solution for both Wireless Information Transfer (WIT) and Wireless Energy Transfer (WET). Specifically, for WET, the additional Line-of-Sight (LoS) path can reduce the effects of large-scale fading caused by blockages and shadows, enhancing the potential of WET. Recent studies have proposed a variety of techniques to optimize the performance of SWIPT/WET systems with the assistance of IRS. For example, a low-complexity alternating optimization algorithm was proposed in \cite{IRSassistedSWIPT}, which meets the secrecy rate requirements while increasing harvested energy by nearly twofold. Other works have jointly optimized the transmit precoder and passive phase shift matrix of the IRS, demonstrating the advantages of using IRS in SWIPT/WET systems \cite{IRSassistedSWIPT2, IRS_WET}. Furthermore, the effectiveness of employing multiple IRSs to improve SWIPT system performance has been demonstrated in \cite{IRSassistedSWIPT3}. These results highlight the potential benefits of using IRS in future wireless communication systems.

    The application of IRS in WET faces a significant challenge in the acquisition of Channel State Information (CSI) between the IRS and its serving PBs/users\cite{wqq_survey, intro_overviewbruno1,intro_overviewbruno2,SWIPT3,intro_overviewonel1,intro_Mon}. The lack of signal processing capability in IRS complicates the estimation of cascaded channels of PBs-to-IRS-to-users. Furthermore, the pilot overhead is proportional to the number of IRS elements and users\cite{pilotoverhead1, pilotoverhead2}, making channel estimation impractical. The absence of active RF links in low-cost reflecting elements also makes pilot transmission for channel estimation impossible. Recent research efforts have explored the problem of IRS channel estimation. Deep learning and compressed sensing-based methods with randomly distributed active sensors have been proposed in \cite{taha2021enabling} to estimate the channel with negligible pilot overhead. Additionally, a power sensor added behind the IRS element can be used to observe the interference phenomenon and superpose the signal in the same phase at the receiver, as described in \cite{zhu2022sensing}. Finally, \cite{ren2021configuring,QintaoCSIfree} propose a technique that collects a large number of actual observations to determine the optimal precoder and IRS phase shift. These innovative approaches offer promising solutions for channel estimation in WET with IRS, which will be critical for the future development of 6G wireless communication systems.

    In a nutshell, the existing works on IRS-assisted WET systems have the following drawbacks:
    \begin{itemize}
        \item The pilot overhead of IRS-assisted WET systems is intolerable. Although recent research attempts to propose a reasonable solution, additional active RF links, hardware, or observation time are required.
        \item Most works concentrated on IRS-assisted WET systems for a small number of energy receivers (ERs). In fact, massive ERs will pose new challenges to channel estimation and IRS-assisted WET systems in the coming IoT era.
        \item Most works assume that the IRS hardware is perfect, which means that the passive phase shift and reflection amplitude can be adjusted independently. Unfortunately, this is not feasible at the current industrial level.
    \end{itemize}

    This paper presents a novel scheme for IRS-assisted WET that does not require CSI (it is also referred to as a CSI-free scheme in this paper) and can accommodate both perfect and imperfect IRS hardware. Our contributions are summarized as follows:
    \begin{itemize}
        \item A CSI-free scheme is proposed. Combined with the proposed simple but effective rotation scheme, the proposed CSI-free scheme can cover the entire space. It is worth noting that our scheme does not require any pilot overhead for the IRS-to-ERs channel, so it is ideal for large-scale IRS elements and massive ERs.
        \item We considered the scenario of massive ERs to meet the massive energy supply demand that may arise in the future. Our scheme outperforms the CSI-based counterpart in scenarios involving large-scale ERs. Furthermore, no re-optimization is required for any ER entry to or removal from this system, lowering maintenance costs significantly.
        \item We propose an extension of the CSI-free scheme from perfect to imperfect hardware and provide a comprehensive mathematical derivation and explanation for the extension. We present different schemes for non-LoS (NLoS) and LoS channels, and provide rigorous proofs and numerical results. Our proposed scheme retains the passive and low-cost nature of the IRS hardware, and is highly compatible with the current hardware model and system architecture. Furthermore, our scheme does not require any modifications to the existing IRS hardware, allowing it to inherit the majority of existing IRS research and ensuring that it is easy to implement.
        \item Numerous experiments are designed. We conducted a thorough examination of the proposed CSI-free scheme's performance in detail. Numerical results illustrate that in massive ERs and large-scale IRS scenarios, such as cases when the number of ERs in the energy coverage range reaches 64 or the number of IRS elements reaches 169, the proposed CSI-free scheme outperforms the CSI-based counterpart by approximately 2 dB. This outcome highlights the feasibility and rationality of the CSI-free scheme in the context of WET assisted by IRS.
    \end{itemize}

    The remainder of this paper is organized as follows. Section \ref{sec:2} provides an overview of the system model, while Section \ref{sec:3} presents a novel CSI-free scheme for both uncoupled and coupled reflection amplitude and phase shift IRS models. The numerical results are presented in Section \ref{sec:4}, and Section \ref{sec:5} summarizes the findings and conclusions of this study.

    \emph{Notation:} $\mathbf{I}_M$ and $\mathbf{1}_M$ denote the $M$ dimension identity matrix and the column vector with all one. $[\cdot]_i$ and $[\cdot]_{i,j}$ denote the $i$-th element of vector and $(i,j)$-th element of matrix, respectively. $\mathbbm{i}=\sqrt{-1}$ is the imaginary unit. $||\cdot||$ and $|\cdot|$ denote the Euclidean norm and absolute value. $\text{diag}(\cdot)$ denotes the diagonal matrix. $(\cdot)^\mathrm{T}$, $(\cdot)^\mathrm{\dagger}$, $(\cdot)^\mathrm{H}$ denote the transpose, conjugate, conjugate transpose operators, respectively. $\Re(\cdot)$ is the real-value operator, while $\Im(\cdot)$ is the imaginary-value operator. $\mathbb{E}(\cdot)$ and $\mathbb{D}(\cdot)$ denote the mathematical expectation and variance, respectively. $Z\sim\mathcal{X}^2(a, b)$ denotes the non-central chi-square distribution with freedom degree $a$ and parameter $b$, and the mean value of $Z$ is $\mathbb{E}(Z)=a+b$ and the variance of $Z$ is $\mathbb{D}(Z)=2(a+2b)$. $\mathcal{U}$, $\mathcal{CN}$ and $\mathcal{N}$ denote the uniform distribution,  circularly symmetric complex Gaussian distribution and Gaussian distribution, respectively.
    \begin{figure}
        \centering
        \includegraphics[width=0.9\linewidth]{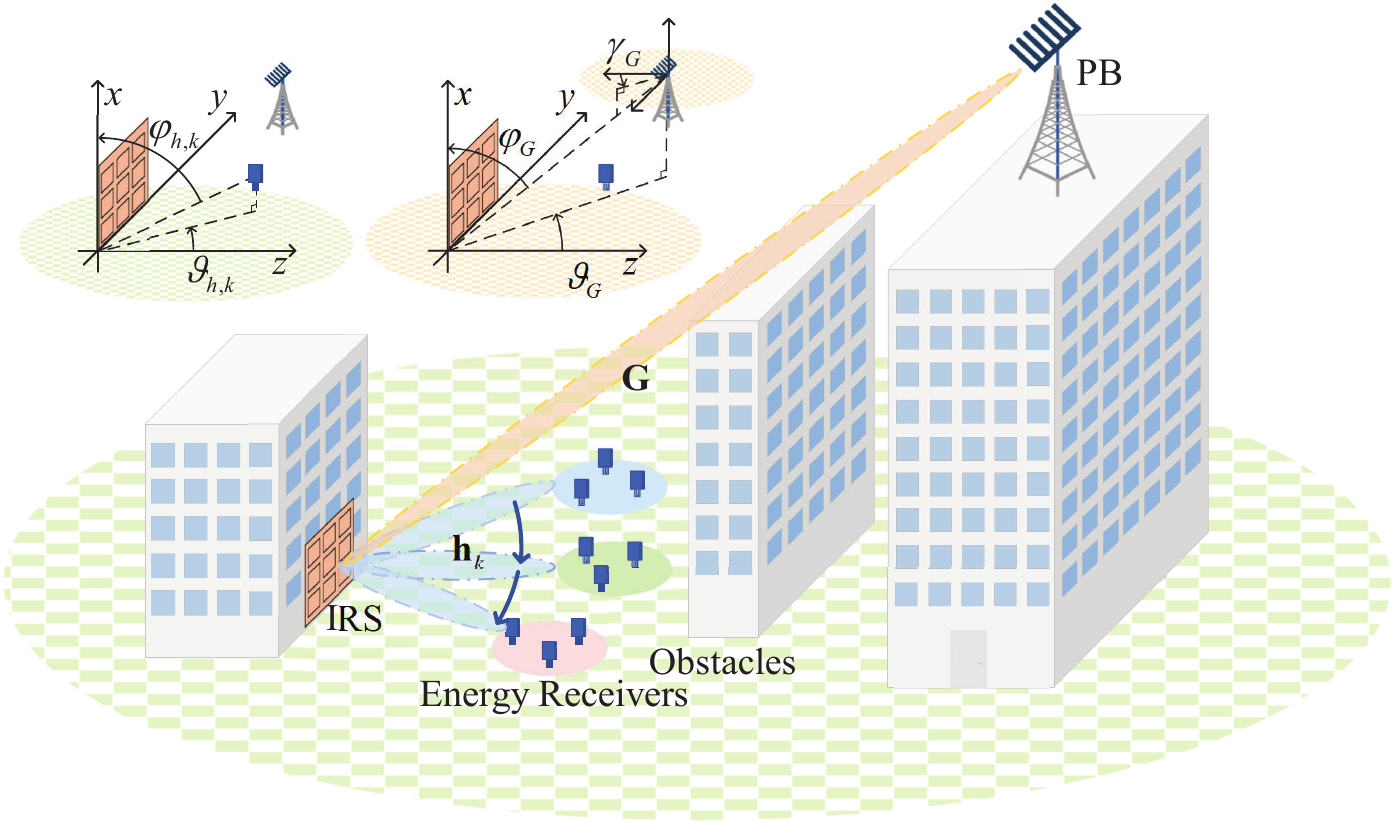}
        \setlength{\abovecaptionskip}{0pt}
        \setlength{\belowcaptionskip}{0pt} 
        \caption{An IRS-assisted WET system.}
        \label{fig:pipeline}
        
    \end{figure}
    \section{System Model}\label{sec:2}
    In this paper, we focus on the Multiple-Input-Single-Output (MISO) IRS-assisted WET downlink scenario, as depicted in Fig. \ref{fig:pipeline}. In this scenario, a PB equipped with a uniform linear array (ULA) with size of $M$ supports WET to massive single-antenna ERs, by employing an IRS with $N_x\times N_y = N$ reflecting elements. Additionally, there is no direct link available due to blockage. 
	
    \subsection{Channel Model}
    Quasi-static flat fading channels are assumed. Specifically, The channel from PB-to-IRS denoted as $\mathbf{G}\in\mathbb{C}^{N\times M}$ is perfectly known, and the channel from the IRS to the $k$-th ER, denoted as $\mathbf{h}_k\in\mathbb{C}^{N\times 1}$, is unknown. We adopt a LoS channel model for $\mathbf{G}$, which can be readily obtained through the channel estimation of the PB and the IRS controller\footnote{\textcolor{red}{To achieve analytical tractability, we employ a LoS channel model for $\mathbf{G}$ and consider it as known, which is reasonable since an IRS can typically provide an additional LoS path and the positions of both the IRS and the PB are usually fixed in practical scenarios.}}\footnote{\textcolor{red}{Many methods can be employed to estimate the channel $\mathbf{G}$, such as location-based \cite{locinfo} and angle-based schemes \cite{angleinfo}, etc.}}\cite{pilotoverhead1}, and a Rician channel model for $\mathbf{h}_k$. Therefore
    \begin{align}
        \mathbf{G}=\sqrt{MN}\boldsymbol{\alpha}_{G,r}(\vartheta_{G},\varphi_{G})\boldsymbol{\alpha}_{G,t}(\gamma_{G})^\mathrm{H},
    \label{eqn:18}
    \end{align}
    where $\varphi_{G}$ $(\vartheta_{G})$ and $\gamma_{G}$ represent the azimuth (elevation) angle of arrival (AoA) and the angle of departure (AoD) from PB to IRS, respectively (shown in Fig. \ref{fig:pipeline} upper left). Since the IRS is an $N_x\times N_y$ uniform planar array (UPA), we have
    \begin{align}
        \boldsymbol{\alpha}_{G,r}(\vartheta_{G},\varphi_{G})=\boldsymbol{\alpha}_{G,x}(u_{G})\otimes\boldsymbol{\alpha}_{G,y}(v_{G}),
    \end{align}
    where $\otimes$ stands for the Kronecker product. $u_{G}\triangleq 2\pi d\cos\varphi_{G} /\lambda=\pi\cos\varphi_{G}$ by setting $d/\lambda=1/2$ without sacrificing generality, where $d$ and $\lambda$ are the element spacing and carrier wavelength. Similarly, we have $v_{G}\triangleq \pi\sin\varphi_{G}\sin\vartheta_{G}$, $z_{G}\triangleq \pi\sin\gamma_{G}$. Hence $\boldsymbol{\alpha}_{G,t}(\gamma_{G})=1/\sqrt{M}\left[1,e^{\mathbbm{i}z_{G}},\cdots,e^{(M-1)\mathbbm{i}z_{G}}\right]^\mathrm{T}=\boldsymbol{\alpha}_{G,t}(z_{G})$, $\boldsymbol{\alpha}_{G,x}(u_{G})=1/\sqrt{N_x}\left[1,e^{\mathbbm{i}u_{G}},\cdots,e^{(N_x-1)\mathbbm{i}u_{G}}\right]^\mathrm{T}$, and $\boldsymbol{\alpha}_{G, y}(v_G)=1/\sqrt{N_y}\left[1,e^{\mathbbm{i}v_G},\cdots,e^{(N_y-1)\mathbbm{i}v_G}\right]^\mathrm{T}$. \textcolor{red}{The channel $\mathbf{h}_k$ from the IRS to $k$-th ER is expressed as
    \begin{align}
        \mathbf{h}_k=&\sqrt{\frac{\kappa_k}{1+\kappa_k}}\textbf{h}_{\text{los}}+\sqrt{\frac{1}{1+\kappa_k}}\textbf{h}_{\text{nlos}} \nonumber\\
        &\sim \sqrt{\frac{1}{1+\kappa_k}}\mathcal{CN}\left(\sqrt{\kappa_k}\left[e^{\mathbbm{i}\Phi_{k,1}}, e^{\mathbbm{i}\Phi_{k,2}},\cdots,e^{\mathbbm{i}\Phi_{k,N}}\right]^\mathrm{T}, \mathbf{I}_N\right),
        \label{eqn:channelhk}
    \end{align}
    where $\kappa_k$ denotes the Rician factor of channel $\mathbf{h}_k$, $\mathbf{h}_\mathrm{los}$ represents the LoS component and $\mathbf{h}_\mathrm{nlos}$ represents the scattering component of $\mathbf{h}_k$.} The $i$-th phase of the array response vector $\boldsymbol{\alpha}_{h,k,r}(\vartheta_{h,k},\varphi_{h,k})=\boldsymbol{\alpha}_{h,k,x}(u_{h,k})\otimes\boldsymbol{\alpha}_{h,k,y}(v_{h,k})$ is represented by $\Phi_{k,i}$ (i.e., $\Phi_{k,i}=\text{arg}([\boldsymbol{\alpha}_{h,k,r}(\vartheta_{h,k},\varphi_{h,k})]_i)$), where the azimuth and elevation angles of departure (AoD) from the IRS to the $k$-th ER are denoted by $\varphi_{h,k}$ and $\vartheta_{h,k}$, respectively. The variables $u_{h,k}$ and $v_{h,k}$ are defined analogously to those in the channel matrix $\mathbf{G}$. For the purpose of simplification and without loss of generality, it is assumed that the IRS and ERs reside in the same horizontal plane, implying $\varphi_{h,k}=\pi/2$. This assumption enables the reduction of the ER's positional model from a three-dimensional to a two-dimensional representation. It should be noted that extending the model from a two-dimensional to a three-dimensional configuration is a straightforward process.

    It is crucial to highlight that the channel $\mathbf{h}_k$ remains entirely unknown. To elaborate, our approach lacks knowledge concerning the phase $\Phi_{k,i}, \forall i\in N$, the value of NLoS component, and the Rician factor $\kappa_k$ present in Eq. \eqref{eqn:channelhk}. Contrarily, the channel $\mathbf{G}$ is fully comprehended and concurrently shared among all ERs.

    \subsection{Practical Phase Shift IRS Model} \label{sec:coupledmodel}
    The investigation of the IRS in both WIT and WET has been extensively conducted. However, numerous studies have overlooked the imperfect hardware properties inherent to IRS systems. Recent research has delved into the impact of these imperfections, modeling each reflecting element as a resonant circuit characterized by specific inductance, capacitance, and resistance values \cite{Rzhang_practicalModel, wqq_survey}. This relationship can be represented as follows:
    \begin{align}
    \beta_i=\beta\left(\theta_{i}\right)=\left(1-\beta_{\text{min} }\right)\left(\frac{\sin \left(\theta_{i}-\eta\right)+1}{2}\right)^{\alpha}+\beta_{\text{min}},\label{eqn:coupled_beta}
    \end{align}
    \textcolor{red}{where $\beta_{\text{min}}$ represents the minimum amplitude, $\eta$ denotes the horizontal distance between $-\pi/2$ and $\beta_{\text{min}}$, and $\alpha$ governs the steepness of the function curve. In practice, IRS circuits are fixed once they are fabricated, making these parameters readily available and easily determined using standard curve fitting tools.} It can be readily deduced that $\beta_i=1,\forall i\in N$ when $\theta_i=\eta+\pi/2,\forall i\in N$. Building upon this model, it has been established that the amplitude response of the reflecting element exhibits a non-linear relationship with its phase shift, precluding independent adjustments. This model will serve as the foundation for the development of our CSI-free scheme in the subsequent sections of this paper.
    \section{IRS assisted massive WET without CSI}\label{sec:3}
    \textcolor{red}{This section will commence with a discussion of uncoupled reflection amplitude and phase shift IRS model, which we will hereafter refer to as the Ideal IRS. This will serve as a foundation for exploring the concept of IRS-assisted WET without knowledge of $\mathbf{h}_k$. The primary objective of this exploration is to enable an investigation into the CSI-free approach for the coupled reflection amplitude and phase shift IRS model, subsequently referred to as the Practical IRS. We will focus solely on the $k$-th ER scenario, as the performance of an individual ER serves as an indicator for the entire system. For the sake of clarity, we will omit the subscript of $\kappa_k$ and assume that all ERs experience a similar channel condition (i.e., $\kappa_k=\kappa,\forall k$) without loss of generality.}

    \subsection{Ideal IRS assisted WET}\label{sec:IdeaIRS}
    With the aid of IRS, the signal received by the $k$-th ER can be expressed as\footnote{The signal received by the $k$-th ER is the sum of $N$ IRS elements' reflected signals. And for WET, the noise impact can be ignored, as widely adopted in the literature\cite{ignoreNoisepower1,ignoreNoisepower2,zeng2014optimized}.}
    \begin{align}
        y_k=&\sqrt{\alpha_{{G}}\alpha_{{h,k}}}\sum\limits_{i=1}^N\mathbf{h}_k^\mathrm{H}\boldsymbol{\Pi}\mathbf{G}\mathbf{w}_ix_k\nonumber\\
        =&\sqrt{\alpha_{{G}}\alpha_{{h,k}}}\sum\limits_{j=1}^N\mathcal{A}_{j}[\mathbf{h}_k]_j[\boldsymbol{\Pi}]_{j,j}x_k,
    \end{align}
    where $\alpha_{{G}}$ and $\alpha_{{h, k}}$ denote the path loss from PB-to-IRS and IRS-to-ER, respectively, $\mathbf{w}_i$ denotes the precoder for $i$-th IRS element. \textcolor{red}{We attempt to design individual beams for each element of the IRS to attain optimal flexibility in manipulating the reflected signals.} $x_k$ is the normalized energy signal, i.e., $\mathbb{E}[x_k^\mathrm{H}x_k]=1$. $\mathcal{A}_j=\mathbf{G}_{j,:}\sum_{i=1}^N\mathbf{w}_i$ represents the signal received by the $j$-th element, where $\mathbf{G}_{j,:}$ represents the $j$-th row of matrix $\mathbf{G}$. Moreover, $\boldsymbol{\Pi}=\mathrm{diag}(e^{\mathbbm{i}\theta_1},\cdots,e^{\mathbbm{i}\theta_N})\mathrm{diag}(\beta_1,\cdots,\beta_N)=\Theta\boldsymbol{\beta}$ is the diagonal phase shift matrix of IRS, where $\theta_i\in[-\pi,\pi]$ and $\beta_i\in[0,1]$ are the phase shift and amplitude reflection coefficient, respectively.
    \begin{lemma}\label{lemma:1}
        Under the LoS PB-to-IRS channel model considered in this paper, the signal incident on the IRS element has the same amplitude but a different phase.
    \end{lemma}
    \begin{IEEEproof}
        Please refer to Appendix \ref{app:A} for detailed proof.
    \end{IEEEproof}

    \begin{Rem}
        According to Lemma \ref{lemma:1}, we know that the amplitude of incident signal is associated to the component $\boldsymbol{\alpha}_{G,t}(z)^\mathrm{H}\sum_{i=1}^N\mathbf{w}_i$. It can be verified that the maximum-ratio transmission (MRT) is the optimal transmit precoder for all IRS elements to maximize incident signal power\cite{tse2005fundamentalsMRT, wu2019intelligent}, and we only need one precoder pointing to IRS for energy transfer from PB to IRS, i.e., $\mathbf{w}_{\mathrm{MRT}}=\sum_{i=1}^N\mathbf{w}_i=\sqrt{P}\frac{\boldsymbol{\alpha}_{G,t}(z)}{||\boldsymbol{\alpha}_{G,t}(z)||}$, where $P$ is the total transmit power. In a nutshell, we have $\mathcal{A}_j=\sqrt{P_e}e^{\mathbbm{i}\mu_j},\forall i\in N$, where $\mu_j$ and $P_e$ represent the different phase and the same power incident on the $j$-th element, respectively.
        \label{remark:1}
    \end{Rem}

    \textcolor{red}{It is important to emphasize that in a more generalized channel model, such as the Rician channel, strict adherence to the characteristics specified in Lemma \ref{lemma:1} is crucial. Given the absence of CSI between the IRS and the ERs, it is vital to ensure uniformity in the incident power directed at each IRS element. Such uniformity ensures equitable performance, particularly when the specifics of the IRS-to-ER channel are undetermined. For environments typified by Rician channels, this equity might necessitate the deployment of supplementary antennas at the PB. The essence of our methodology in formulating distinct beams for every IRS element is rooted in this foundational concept.
    }

    Then the energy received by the $k$-th ER can be expressed as
    \begin{align}
        E_k &= \alpha_{{G}}\alpha_{{h,k}}\left|\sum\limits_{j=1}^N\mathcal{A}_j[\mathbf{h}_k]_j[\boldsymbol{\Pi}]_{j,j}\right|^2\nonumber\\
        &\overset{(a)}=\alpha_{{G}}\alpha_{{h,k}}P_e\left|\mathbf{1}_N^\mathrm{T}\boldsymbol{\Pi}\boldsymbol{\mu}\mathbf{h}_k\right|^2\nonumber\\
        &\overset{(b)}=\alpha_{{G}}\alpha_{{h,k}}P_e\left|\mathbf{1}_N^\mathrm{T}\mathbf{h}_k^\S \right|^2\label{eqn:receiveEk},
    \end{align}
    where $(a)$ comes from Lemma 1 and Remark 1 where $\boldsymbol{\mu}=\mathrm{diag}(e^{\mathbbm{i}\mu_1},\cdots,e^{\mathbbm{i}\mu_N})$ and $(b)$ comes from the cascade channel representation as $\mathbf{h}^\S_k=\boldsymbol{\Pi}\boldsymbol{\mu}\mathbf{h}_k$.

    For the Ideal IRS model, we set $|\beta_j|=1,\forall j\in N$ to maximize the reflection signal power, and $\mathbf{\Pi}$ can be simplified as $\Theta=\mathrm{diag}(e^{\mathbbm{i}\theta_1},\cdots,e^{\mathbbm{i}\theta_N})$. Then we can obtain
    \begin{align}
        &\mathbf{h}_k^{\S}=\Theta\boldsymbol{\mu}\mathbf{h}_k=\sqrt{\frac{\kappa}{1+\kappa}}\Theta\boldsymbol{\mu}\textbf{h}_{\text{los}}+\sqrt{\frac{1}{1+\kappa}}\Theta\boldsymbol{\mu}\textbf{h}_{\text{nlos}}\nonumber \\
        &\sim \sqrt{\frac{1}{1+\kappa}}\mathcal{CN}\left(\sqrt{\kappa}\left[e^{\mathbbm{i}(\Phi_{k,1}+\theta_1+\mu_1)},\cdots,e^{\mathbbm{i}(\Phi_{k,N}+\theta_{N}+\mu_N)}\right]^\mathrm{T}, \mathbf{R}^\S\right),
        \label{eqn:cascadehIdeaIRS}
    \end{align}
    where $\mathbf{R}^\S=\Theta\boldsymbol{\mu}\mathbf{I}_N\boldsymbol{\mu}^\mathrm{H}\Theta^\mathrm{H}=\mathbf{I}_N$. Therefore Eq. \eqref{eqn:receiveEk} can be reconstructed as
    \begin{align}
        E_k = P_e\left|\mathbf{1}_N^\mathrm{T}\mathbf{h}_k^{\S}\right|^2=P_e\left(\Re{(\mathbf{1}_N^\mathrm{T}\mathbf{h}_k^{\S})}^2+\Im{(\mathbf{1}_N^\mathrm{T}\mathbf{h}_k^{\S})}^2\right),
        \label{eqn:EkwithoutPathloss}
    \end{align}
    and the component of the path loss $\alpha_G\alpha_{h,k}$ is ignored\footnote{Note that Eq. \eqref{eqn:EkwithoutPathloss} is a  scaled representation of the actual harvested energy, with path loss being disregarded. Consequently, the results derived from this equation exhibit analogous trends.}.

    It is essential to reiterate that the channel $\mathbf{h}_k$ remains entirely unknown. Nevertheless, we know the incident phase $\mu_i,\forall i\in N$ and IRS phase shift $\theta_i,\forall i\in N$ in cascaded channel $\mathbf{h}^\S_k$, since PB-to-IRS channel is perfectly known and $\theta_i,\forall i\in N$ is the phase shift we need to adjust. Furthermore, we know the variance matrix $\mathbf{R}^\S=\mathbf{I}_N$, which forces us to start with its statistical value. We have
    \begin{align}
        \Re{(\mathbf{1}_N^\mathrm{T}\mathbf{h}_k^{\S})}=&\sum_{j=1}^N\Re{\left(\sqrt{\frac{\kappa}{1+\kappa}}e^{\mathbbm{i}(\Phi_{k,j}+\theta_j+\mu_j)}+\left[\sqrt{\frac{1}{1+\kappa}}\mathcal{CN}(\mathbf{0}, \mathbf{R}^\S)\right]_j\right)}\nonumber\\
        &\sim \sqrt{\frac{1}{2(\kappa+1)}}\mathcal{N}(\sqrt{2\kappa}u,R_{\Sigma}).\label{eqn:realpart}
    \end{align}
    Similarly, we have
    \begin{align}
        \Im{(\mathbf{1}_N^\mathrm{T}\mathbf{h}_k^{\S})}\sim \sqrt{\frac{1}{2(\kappa+1)}}\mathcal{N}(\sqrt{2\kappa}v,R_{\Sigma}) \label{eqn:imagpart},
    \end{align}
    where $u=\sum_{i=1}^N\cos(\Phi_{k,i}+\theta_i+\mu_i)$, $v=\sum_{i=1}^N\sin(\Phi_{k,i}+\theta_i+\mu_i)$, and $R_{\Sigma}=\mathbf{1}_N\mathbf{I}_N\mathbf{1}_N^\mathrm{T}=N$.

    Note that Eq. \eqref{eqn:realpart} and Eq. \eqref{eqn:imagpart} have the same variance but different mean values, thus square sum term illustrated in Eq. \eqref{eqn:EkwithoutPathloss} should follow the non-central chi-square distribution with two degrees of freedom\cite{lopez2020csi}, yielding
    \begin{align}
        E_k\sim \frac{P_eR_{\Sigma}}{2(\kappa+1)}\mathcal{X}^2\left(2,\frac{2\kappa (u^2+v^2)}{R_{\Sigma}}\right)
        \label{eqn:noncentralChi-sqrt},
    \end{align}
    and the mean value of $E_k$ is
    \begin{align}
        &\mathbb{E}\left[E_k\right]=\frac{P_eR_{\Sigma}}{2(\kappa+1)}\left(2+\frac{2\kappa(u^2+v^2)}{R_{\Sigma}}\right) \label{eqn:expressionMean}.
    \end{align}

    It is clear that maximizing $\mathbb{E}[E_k]$ is equivalent to maximizing $(u^2 + v^2)$, given that $P_e$ can be confirmed by MRT as illustrated in Lemma \ref{lemma:1} and Remark \ref{remark:1}, and $R_\Sigma=N$ holds for Ideal IRS. For the convenience of subsequent representation in this paper, we define $E_{eq}=u^2+v^2$ as the \emph{Equivalent Received Energy}. Furthermore, the Rician factor $\kappa$ exists objectively and also has no impact on Ideal IRS's $E_{eq}$ maximization. It is worth noting that we do not care about the incident phase (i.e., $\mu_j,j\in N$) in this case, since the phase of IRS can be freely adjusted to align it. Then we can derive the maximum value as $E_{eq}=u^2+v^2=N^2$ when the phase shift of IRS $\theta_i$ satisfies
    \begin{align}
        \Phi_{k,i}+\theta_i+\mu_i=\Phi_{k,j}+\theta_j+\mu_j,\forall i,j\in N.
        \label{eqn:ideaIRStheta}
    \end{align}

    Eq. \eqref{eqn:ideaIRStheta} shows that for the Ideal IRS, aligning the incident phase (i.e., $\mu_j, \forall j\in N$) from the PB-to-IRS and matching the reflection phase required for the given beam direction $\vartheta_{h,k}$ (i.e., $\Phi_{k,j}=-(\text{mod}(j,N_y)-1)\pi\sin\vartheta_{h,k}, \forall j\in N$ since $\varphi_{h,k}=\pi/2$) is sufficient to maximize $\mathbb{E}(E_k)$.

    However, it is unfortunate that the location of the $k$-th ER (i.e., $\vartheta_{h,k}$) is unknown. This issue will be addressed in the subsequent section.

    \subsection{Beam rotation scheme for Ideal IRS}\label{sec:beamrotary_perfect}
    Without knowing the location of the $k$-th ER (i.e., $\vartheta_{h,k}$), we can assume the beam is pointing to the direction $\hat{\vartheta}_{h,v}$ and maximizes the average energy in $\hat{\vartheta}_{h,v}$ using Eq. \eqref{eqn:ideaIRStheta}. It bears repeating that the pointing direction of the beam is not necessarily identical to the $k$-th ER's actual location. Then we will rotate the beam's direction to accommodate all possible ERs in a single period.

    In order to select a reasonable rotation scheme, we first describe the relationship between the antenna pattern and the direction $\hat{\vartheta}_{h,v}$ as\cite{Beamwidth} 
    \begin{align}
        {F}_{\hat{\vartheta}_{h,v}}(\omega)=\frac{\sin({\frac{N\pi d}{\lambda}(\sin(\omega)-\sin(\hat{\vartheta}_{h,v}))})}{\sin({\frac{\pi d}{\lambda}(\sin(\omega)-\sin(\hat{\vartheta}_{h,v}))})}.
        \label{eqn:antennapattern}
    \end{align}

\begin{figure}
        \centering
        \includegraphics[width=0.9\linewidth]{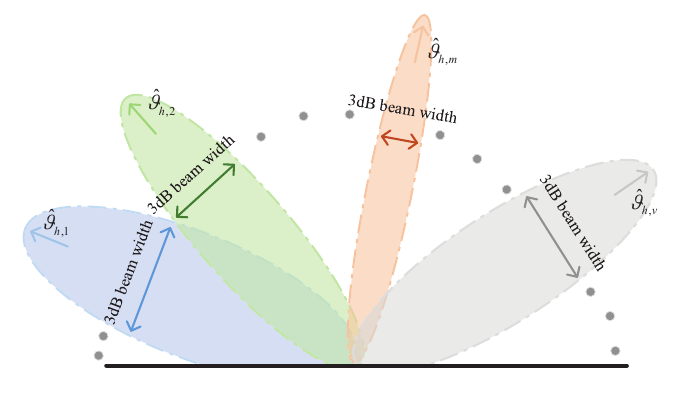}
        \setlength{\abovecaptionskip}{0pt}
        \setlength{\belowcaptionskip}{0pt} 
        \caption{A beam rotation scheme that achieves comprehensive coverage in all directions with the minimum number of beams.}
        \label{fig:Beamwidth}
    \end{figure}

    Eq. \eqref{eqn:antennapattern} depicts the beam gain at various angles $\omega$ when the direction $\hat{\vartheta}_{h,v}$ is determined. Then a reasonable beam rotation scheme can be illustrated in Fig. \ref{fig:Beamwidth}, which can be obtained through selecting an initial phase (e.g., $\hat{\vartheta}_{h,1}=\pi/2$), calculating the beam width by setting ${F}_{\hat{\vartheta}_{h,1}}(\omega)=1/\sqrt{2}$ and recursion\footnote{In practical, it is possible to set ${F}_{\hat{\vartheta}_{h,1}}(\omega)=1/\sqrt{2}\pm \delta$, where $[-\delta, \delta]$ represents a small feasible interval, thus allowing the beam to cover the entire space more evenly.}, ensuring that the minimum number of beams is employed and can be covered by the 3 dB beam gain in any direction. By utilizing this beam rotation scheme, the spatial range $[-\pi/2, \pi/2]$ can be covered in a single rotation period in order to achieve the CSI-free WET scheme.

    Let $\mathbf{P}_{\hat{\vartheta}_{h,v}}$ represent the energy coverage when the pointing direction is $\hat{\vartheta}_{h,v}$, Consequently, the complete energy coverage (also referred to as a heat map) within a single rotation period can be denoted by
    \begin{align}
        \mathbf{H}_{cov}=\frac{1}{V}\sum\limits^V_{v=1}\mathbf{P}_{\hat{\vartheta}_{h,v}},\label{eqn:heatmapExpression}
    \end{align}
    where $V$ signifies the minimal number of beams necessary to cover the entire range from $-\pi/2$ to $\pi/2$, while $1/V$ represents the normalization average time for each $\mathbf{P}_{\hat{\vartheta}_{h,v}}$. Further optimization can be pursued based on supplementary information, such as location and distance\cite{OneLRotaryAntenna, ke2022power,Xiang2023RobustNO}, which will be reserved for future work.

    It is worth noting that the antenna pattern does not directly correspond to the actual energy coverage, as various factors, such as large-scale or small-scale fading, come into play. These effects, unknown in the proposed scheme, prevent the full consideration of their impact. Nonetheless, when solely accounting for distance-dependent path loss, a larger beam gain can transmit energy over greater distances, thereby allowing the beam gain to be regarded as proportional to the energy coverage. In summary, in the context of an unknown channel, it can only be ensured that each direction possesses a specific reflection beam gain.

    \subsection{Practical IRS assisted WET}
    In this section, we consider a Practical IRS-assisted WET, with $\boldsymbol{\Pi}=\Theta\boldsymbol{\beta}$. Then we can reformulate  Eq. \eqref{eqn:cascadehIdeaIRS} as
    \begin{align}
        &\mathbf{h}_k^{\S}=\Theta\boldsymbol{\mu}\boldsymbol{\beta}\mathbf{h}_k=\sqrt{\frac{\kappa}{1+\kappa}}\Theta\boldsymbol{\mu}\boldsymbol{\beta}\textbf{h}_{\text{los}}+\sqrt{\frac{1}{1+\kappa}}\Theta\boldsymbol{\mu}\boldsymbol{\beta}\textbf{h}_{\text{nlos}}\nonumber\\
        &\sim \sqrt{\frac{1}{1+\kappa}}\mathcal{CN}(\sqrt{\kappa}\left[\beta_1e^{\mathbbm{i}(\Phi_{k,1}+\theta_1+\mu_1)},\cdots,\beta_{N}e^{\mathbbm{i}(\Phi_{k,N}+\theta_{N}+\mu_N)}\right]^\mathrm{T}, \mathbf{\bar{R}^\S}),
    \end{align}
    where $\mathbf{\bar{R}}^{\S}=\boldsymbol{\beta}\mathbf{R}^\S\boldsymbol{\beta}^\mathrm{H}$. And the energy $E_k$ still follows the non-central chi-square distribution as depicted in Eq. \eqref{eqn:noncentralChi-sqrt}, but
    \begin{align}
        u=&\sum\limits_{i=1}^N\beta_i\cos(\Phi_{k,i}+\theta_i+\mu_i)\nonumber\\
        =&\sum\limits_{i=1}^N\left(\left(1-\beta_{\min }\right)\left(\frac{\sin \left(\theta_{i}-\eta\right)+1}{2}\right)^{\alpha}+\beta_{\min }\right)\cos(\Phi_{k,i}+\theta_{i}+\mu_i),\\
        v=&\sum\limits_{i=1}^N\beta_i\sin(\Phi_{k,i}+\theta_i+\mu_i)\nonumber\\
        =&\sum\limits_{i=1}^N\left(\left(1-\beta_{\min }\right)\left(\frac{\sin \left(\theta_{i}-\eta\right)+1}{2}\right)^{\alpha}+\beta_{\min }\right)\sin(\Phi_{k,i}+\theta_{i}+\mu_i),
    \end{align}
    and $R_{\Sigma}^\S=\sum_{i=1}^N\beta_i^2$ according to Eq. \eqref{eqn:coupled_beta} and Eq. \eqref{eqn:imagpart}. The corresponding optimization problem can then be formulated as
    \begin{align}
        \text{(P1): }\max\limits_{\Theta,\mathbf{w}}&\,\, \frac{P_e}{2(\kappa+1)}\left(2R_{\Sigma}^\S+2\kappa E_{eq}\right)\label{eqn:eqnP1}\\
        \text{s.t.}&\,\, ||\mathbf{w}||^2\leq P,\tag{\ref{eqn:eqnP1}a} \label{eqn:P1constraint1}\\
        &\,\,\left|[\Theta]_i\right|=1, \forall i\in N,\tag{\ref{eqn:eqnP1}b}\label{eqn:P1constraint2}
    \end{align}
    where the objective (P1) is the expectation of harvest energy, $\mathbf{w}$ and $\Theta$ are the transmit precoder of PB and phase shift matrix of IRS. Transmit power and passive phase shift limits cause constraints in \eqref{eqn:P1constraint1} and \eqref{eqn:P1constraint2}, respectively.

    In contrast to the Ideal IRS, phase adjustment in the Practical IRS impacts the reflection amplitude, which implies that the phase cannot be freely adjusted. As a result, maximizing $E_{eq}$ might decrease the value of $R_\Sigma^\S$, and the incident phase (i.e., $\mu_j,j\in N$) could also influence the value of $E_{eq}$. Both factors necessitate a more detailed examination. We start by simplifying the objective (P1) as follows:
    \begin{itemize}
        \item[$\bullet$] $\kappa$ is very small. Under this condition, we obtain $\lim_{\kappa\to 0}\frac{P_e}{2(\kappa+1)}(2R_{\Sigma}^\S+2\kappa E_{eq})=P_eR_{\Sigma}^\S$, 
         indicating that we only need to set $\Theta_i=\eta+\pi/2, \forall i\in N$ (refer to Section \ref{sec:coupledmodel}) to maximize the reflected amplitude (i.e., $|\beta_i|=1,\forall i\in N$). Therefore, we have $\max({P_e R_\Sigma^\S})=P_eN$. These results demonstrate that when $\kappa$ is small, effectively concentrating energy in an unknown channel by adjusting the phase is not possible, and we can only maximize the reflection amplitude to minimize energy reflection loss, resulting in $R_\Sigma^\S = N$. Furthermore, MRT can maximize the incident signal power, which still makes MRT the optimal choice in this scenario. For clarity, the \emph{direct method} of maximizing the reflection amplitude (i.e., setting $|\beta_i|=1,\forall i\in N$) and maximizing the incident signal power (i.e., using MRT as the precoder) is referred to as \emph{DM}.
    \end{itemize}

    \begin{itemize}
        \item[$\bullet$] $\kappa$ is comparatively large. Different from mentioned above, we have
        \begin{align}
            &E_{eq}=u^2+v^2\nonumber\\
            =&\left(\sum\limits_{i=1}^N\beta_i\cos(\Phi_{k,i}+\theta_i+\mu_i)\right)^2+\left(\sum\limits_{j=1}^N\beta_j\sin(\Phi_{k,j}+\theta_j+\mu_j)\right)^2\nonumber\\
            =&\sum\limits_{i=1}^N\beta_i^2+2\sum\limits_{t=1}^{N-1}\sum\limits_{l=t+1}^N\beta_t\beta_l\cos(\Phi_{k,t}+\theta_t+\mu_t-\Phi_{k,l}-\theta_l-\mu_l).\label{eqn:ignoreRsigma}
        \end{align}
        Considering that the number of IRS elements $N$ is sufficiently large (e.g., $N=100$), it can be ensured that $E_{eq} \gg\sum_{i=1}^N\beta_i^2=R_{\Sigma}^\S$, according to Eq. \eqref{eqn:ignoreRsigma}. Consequently, maximizing the average energy $\mathbb{E}[E_k]=\frac{P_e}{2(\kappa+1)}(2R_{\Sigma}^\S+2\kappa E_{eq})$ is equivalent to maximizing $P_eE_{eq}$.  For the sake of clarity, this optimization method, which maximizes $P_e E_{eq}$, is referred to as the \emph{Optimization Method (OM)}.
        \end{itemize}
    
    \textcolor{red}{Note that both the \emph{DM} and \emph{OM} schemes are designed to fine-tune the phase of the IRS elements, aiming to enhance the energy received at the ERs. Given the intrinsic link between the reflection amplitude of the IRS elements and their phase, the \emph{OM} scheme balances the tasks of amplifying the reflection amplitude and synchronizing the signal phase. Conversely, the \emph{DM} approach predominantly concentrates on amplifying the reflection amplitude of the IRS elements, considering that a feeble LoS channel might not derive substantial advantage from meticulous phase alignment.
    }

    Before further discussing the \emph{OM}, we show the boundary of $\kappa$ between \emph{DM} and \emph{OM}. It is worth noting that we only maximize the reflected amplitude in \emph{DM}. Therefore, we have\footnote{It is worth noting that when looking for the boundary of $\kappa$, there is no prerequisite for $\kappa$ to be very small or comparatively large, and recalculating the numerical results using the two methods is necessary.}
    \begin{align}
        E_{eq}=&\sum\limits_{i=1}^N\beta_i^2+2\sum\limits_{t=1}^{N-1}\sum\limits_{l=t+1}^N\beta_t\beta_l\cos(\Phi_{k,t}+\theta_t+\mu_t-\Phi_{k,l}-\theta_l-\mu_l)\nonumber\\
        \overset{(a)}=&N+2\sum\limits_{t=1}^{N-1}\sum\limits_{l=t+1}^N\cos(\Phi_{k,t}+\mu_t-\Phi_{k,l}-\mu_l)\label{eqn:uv2On1}\\
        \overset{(b)}\approx&N,\label{eqn:uv2On}
    \end{align}
    where $(a)$ is derived from adjusting $\theta_i=\eta+\pi/2,\forall i\in N$ in \emph{DM} to obtain $|\beta_i|=1,\forall i\in N$. And $(b)$ follows Lemma \ref{lemma:2} below.
    \begin{lemma}
        The phase $\Phi_{k,t}+\mu_t-\Phi_{k,l}-\mu_l, \forall t=1,\cdots,N-1, \forall l=t+1,\cdots,N$ in Eq \eqref{eqn:uv2On1} can be considered as uniform distribution; thus $2\sum_{t=1}^{N-1}\sum_{l=t+1}^N\cos(\Phi_{k,t}+\mu_t-\Phi_{k,l}-\mu_l)\approx 0$. \label{lemma:2}
    \end{lemma}
    \begin{IEEEproof}
        Please refer to Appendix \ref{app:B} for detailed proof.
    \end{IEEEproof}

    \begin{Rem}
        The intuition behind this Lemma can be explained as follows: In \emph{DM}, we only maximize the amplitude without aligning the phase, resulting in $N\leq E_{eq}\ll \max(E_{eq})=N^2$, and approximately equal to $N$. Lemma 2 indicates the value range of $E_{eq}$ in \emph{DM}, laying the foundation for the subsequent analysis of the $\kappa$ boundary between \emph{OM} and \emph{DM}.
        \label{remark:2}
    \end{Rem}

    \textcolor{red}{Thus, when \emph{DM} outperforms \emph{OM}, we have
    \begin{align}
        \frac{P_e}{2(\kappa+1)}(2R_{\Sigma}^\S+2\kappa E_{eq})\leq \frac{P_e}{2(\kappa+1)}(2N+2\kappa N),\label{eqn:kboundary}
    \end{align}
    and the boundary of $\kappa$ is obtained as $\kappa_{B}= \frac{N-R_\Sigma^\S}{E_{eq}-N}$, which means
    \begin{align}
        \kappa &
        \begin{cases}
            \leq \kappa_B, \text{\emph{DM} outperforms \emph{OM}},\\
            >\kappa_B, \text{\emph{OM} outperforms \emph{DM}}.
        \end{cases} \label{eqn:neq1}
    \end{align}}
    
    \textcolor{red}{Note that the boundary $\kappa_B$ defines an interval for selecting between \emph{DM} and \emph{OM}. When the Rician factor is weaker than $\kappa_B$ (i.e., $\kappa < \kappa_B$), we recommend choosing \emph{DM}. Conversely, \emph{OM} is the preferred choice.}

    Since we have not yet discussed how to calculate $E_{eq}$ in \emph{OM}, we will continue to address this boundary of $\kappa$ later (see Remark \ref{remark:4}).

    Focusing on \emph{OM}, as previously mentioned, when $\kappa$ is comparatively large, problem $\text{(P1)}$ can be equivalently formulated as
    \begin{align}
        \text{(P2): }\max\limits_{\Theta,\mathbf{w}}&\,\, P_eE_{eq}\label{eqn:eqnP2}\\
        \text{s.t.}&\,\,(\ref{eqn:P1constraint1}), (\ref{eqn:P1constraint2})\tag{\ref{eqn:eqnP2}a}.
    \end{align}

    Due to the interdependence of reflection amplitude and phase shift, the IRS phase cannot be adjusted freely. Consequently, the incident phase might influence the maximum value of $E_{eq}$, necessitating a comprehensive investigation.

    \begin{lemma}
        The $\mathbf{w}_{\text{MRT}}$ in Remark \ref{remark:1} remains the optimal precoder for problem $\text{(P2)}$, as any modification to $\mathbf{w}_{\text{MRT}}$ does not lead to an increase (i.e., it may decrease or remain unchanged) in the incident power, and the additional phase introduced by the alteration has no effect on maximizing $E_{eq}$. \label{lemma:3}
    \end{lemma}
    \begin{IEEEproof}
        Please refer to Appendix \ref{app:C} for detailed proof.
    \end{IEEEproof}

    \begin{Rem}
        Lemma \ref{lemma:3} establishes that $\mathbf{w}_\text{MRT}$ is the optimal precoder for $\text{(P2)}$. Consequently, problem $\text{(P2)}$ can be separated into two independent maximization problems. This implies that problem $\text{(P2)}$ is equivalent to first employing $\mathbf{w}_\text{MRT}$ as the precoder with $||\mathbf{w}_{\text{MRT}}||^2=P$, and then maximizing $E_{eq}$ under this condition.
        \label{remark:3}
    \end{Rem}

    Hence, the problem $\text{(P2)}$ can be simplified to
    \begin{align}
        \text{(P2.1): }\max\limits_{\Theta}&\,\,E_{eq}\label{eqn:eqnP2_1}\\
        \text{s.t.}&\,(\ref{eqn:P1constraint2}).\tag{\ref{eqn:eqnP2_1}a}
    \end{align}

    Nonetheless, problem $\text{(P2.1)}$ is non-convex due to the presence of sine and cosine terms. We propose an Alternating Optimization (AO)-based algorithm to approximate the solution for $\text{(P2.1)}$ by optimizing one phase shift of the $N$ elements at a time, while keeping the others constant, until the objective value in $\text{(P2.1)}$ converges.

    \begin{algorithm}[H]
        % \begin{spacing}{1.0}
        % \algsetup{linenosize=\footnotesize} \scriptsize
        % 表单独的行间距设置
        % 
        \caption{AO-based algorithm for solving problem (P2.1).}

        \begin{algorithmic}[1]
            \REQUIRE~\ The $v$-th determined beam direction $\hat{\vartheta}_{h,v}$; PB-to-IRS channel $\mathbf{G}$; The Practical IRS parameters $\beta_\text{min}$, $\eta$ and $\alpha$.

            \ENSURE~\ Optimal $\{\theta^*_n\}^N_{n=1}$.

            \STATE Initialization: $\{\theta_n\}^N_{n=1}\leftarrow \text{\emph{Random initialization}}$
            \STATE Calculate $\mu_j\leftarrow \arg(\mathbf{G}_{j,:}\mathbf{w}_{\mathrm{MRT}}),\forall j\in N$
            \STATE Calculate $\hat{\Phi}_{k,v}\leftarrow -(\text{mod}(j,N_y)-1)\pi\sin\hat{\vartheta}_{h,v},\forall j\in N$\\
            \STATE Calculate $\beta_j,\forall j\in N \leftarrow \text{\emph{Eq. \eqref{eqn:coupled_beta}}}$
            \REPEAT
            \FOR {t=1:N}
                \STATE Find the optimal $\theta_t^*$ as the solution to (P2.1) while other $\{\theta_n\}^N_{n=1,n\neq t}$ are fixed;
                \STATE Update $\theta_t=\theta_t^*$ and $\beta_t$ by Eq.\eqref{eqn:coupled_beta};
            \ENDFOR
            \UNTIL{converges or reaches the maximum times}
        \end{algorithmic}
    % \end{spacing}
    \end{algorithm}
    \begin{Rem}
        \textcolor{red}{
        It is important to emphasize that although the presented AO-based algorithm operates iteratively, the computational overhead it introduces remains within acceptable limits. The necessity for optimization arises solely upon the initial application of the proposed OM/DM scheme, contingent upon determining factors such as the channel between the IRS and PB, the count of IRS elements, and the IRS's coupling parameters. Furthermore, introducing or excluding ERs from this system does not mandate a comprehensive re-optimization, thereby substantially mitigating the associated computational costs.
        }
    \end{Rem}

    \subsection{Beam rotation scheme for Practical IRS}\label{sec:beamrotary_imperfect}
    Finally, it is necessary to reevaluate the rotation scheme for the Practical IRS. Fortunately, the relationship between the antenna pattern and direction, as depicted in Eq. \eqref{eqn:antennapattern}, indicates that the beam width is solely associated with the IRS elements $N$, element spacing $d$, carrier wavelength $\lambda$, and direction $\hat{\vartheta}_{h,v}$. In other words, once the IRS parameters are confirmed, the beam width is only connected to the direction $\hat{\vartheta}_{h,v}$ and is independent of other factors such as incident power and reflecting amplitude. In summary, there is no significant distinction between the beamforming stages of Ideal IRS and Practical IRS, and the beam rotation scheme proposed in Section \ref{sec:IdeaIRS} remains applicable for Practical IRS to cover the entire space without any modifications. With this, the design of our CSI-free schemes for both Ideal IRS and Practical IRS is complete.

    \section{Simulation Results}\label{sec:4}
    In this section, we present numerical results to demonstrate the performance of the proposed CSI-free WET scheme. We consider a system operating at a carrier frequency of 915 MHz, corresponding to a signal attenuation of 31.6 dB at a reference distance of 1 meter, and we set the exponent to 2.2 for the PB-to-IRS channel and 2.7 for the IRS-to-ER channel. The PB's total transmit power is set to $P=1$ W, with $M=4$ antennas featuring a 15 dBi antenna gain. The distance between the PB and the IRS is set to 20 meters. The IRS-assisted energy charging area has a maximum radius of 4 meters, with ERs uniformly distributed within it. Since the IRS reflects signals only in its front half-sphere rather than isotropically, we assume that each reflecting element has a 3 dBi gain\cite{IRS_WET}. We employ a linear energy harvesting model, as used in various literature \cite{lopez2019statistical,zeng2014optimized}, with a conversion efficiency set to 0.45, sensitivity set to -24 dBm, and saturation set to -8 dBm\cite{lopez2020csi}, which represent the conversion efficiency from incident power to harvested power, the minimum and maximum RF input power, respectively. Other required parameters are configured as follows, unless otherwise specified: $\beta_{\text{min}}=0.2$, $\alpha=1.6$, $\eta = 0.43\pi$, and $N=100$.

    It is worth noting that instead of directly discussing energy $E_k$, we extensively utilized the $E_{eq}=u^2+v^2$ expression in \ref{exp:1} and \ref{exp:2}. This is because $E_{eq}$ can be considered as a scaled version of $\mathbb{E}[E_k]$ and has a clear relationship with the number of IRS elements $N$ and the incident phase $\mu_j,\forall j\in N$. This approach allows us to disregard other factors (e.g., the Rician factor $\kappa$) and focus on maximizing the average energy $\mathbb{E}[E_k]$.

    \subsection{On the Impact of Coupling Parameters}\label{exp:1}
\begin{figure}
        \centering
        \includegraphics[width=0.92\linewidth]{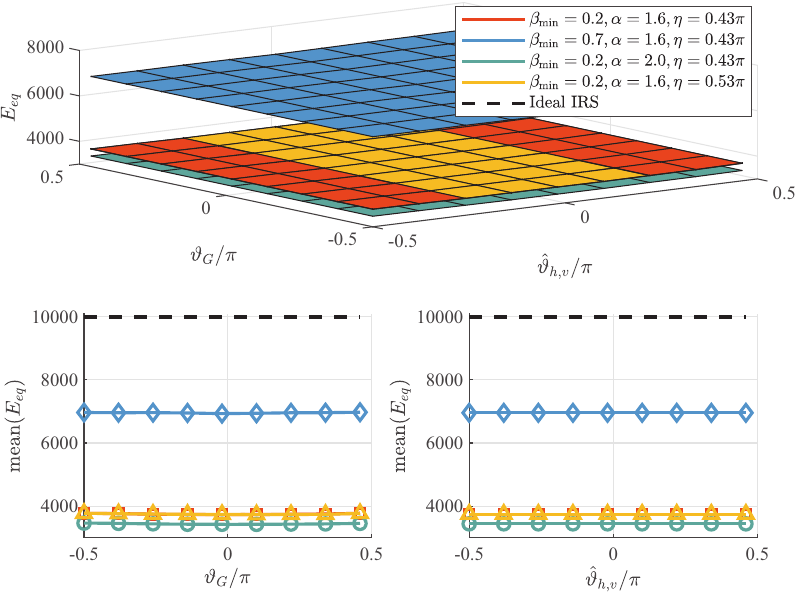}
        \setlength{\abovecaptionskip}{0pt}
        \setlength{\belowcaptionskip}{0pt} 
        \caption{Impact of coupling parameters with various $\vartheta_G$ under proposed CSI-free scheme with $\varphi_G=\pi/4$. The upper figure shows the impact of $\vartheta_G$ and $\hat{\vartheta}_{h,v}$. To better illustrate, we averaged the upper figure's $\hat{\vartheta}_{h,v}$- and $\vartheta_G$-axes, as shown in the right and left bottoms of the figure.}
        \label{fig:exam1}
    \end{figure}

    We first demonstrate the influence of coupling parameters on overall performance. We consider the assumption of $\varphi_G=\pi/4$, and the trend of the \emph{Equivalent Received Energy} is depicted in Fig. \ref{fig:exam1} for different values of $\vartheta_G$ and $\hat{\vartheta}_{h,v}$. \textcolor{red}{Fig. \ref{fig:exam1} reveals that the minimal amplitude $\beta_{\text{min}}$ has a considerable impact on the results, followed by $\alpha$, while the effect of $\eta$ is negligible. This observation is reasonable, as a larger $\beta_{\text{min}}$ (or larger $\alpha$) results in greater fluctuation (or steeper slope) of the IRS reflection amplitude and phase shift coupling curve, consequently leading to more limited phase adjustment. This trade-off between reflection amplitude and phase shift contributes to a significant decline in performance.}

    \textcolor{red}{Additionally, observing Fig. \ref{fig:exam1}, our proposed algorithm consistently performs well across all angles, indicating its effectiveness in focusing energy at different angles while being negligible affected by changes in angle when the circuit parameters are fixed\footnote{An extremely special case occurs when the IRS, PB, and ERs are on the same horizon; the IRS only needs to maximize the reflection amplitude to achieve total reflection. In specific directions, the IRS can be immune to imperfect hardware. Due to space limitations, we will not discuss this case further.}. In subsequent experiments, we use $\varphi_G= \pi / 4 $ and $\vartheta_G = \pi/3 $, unless otherwise specified.}

    \subsection{On the Impact of $N$}\label{exp:2}
\begin{figure}
        \centering
        \includegraphics[width=0.9\linewidth]{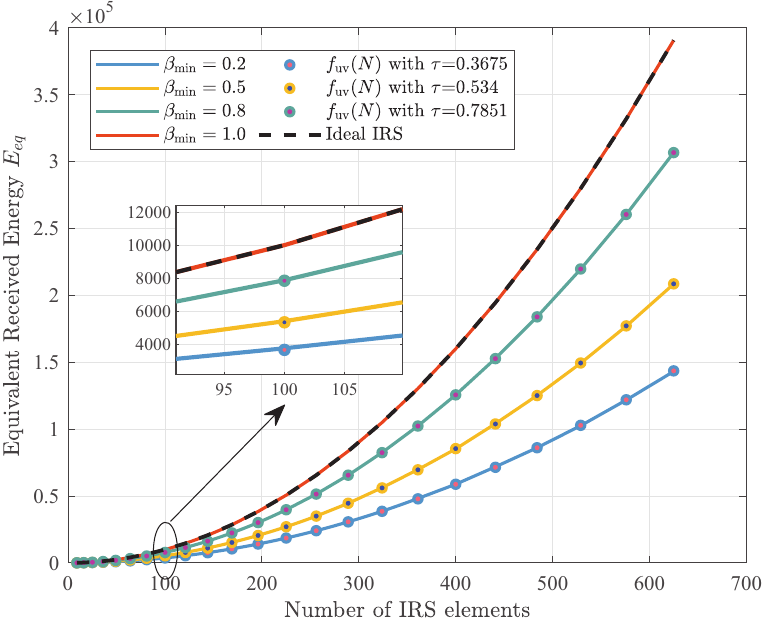}
        \setlength{\abovecaptionskip}{0pt}
        \setlength{\belowcaptionskip}{0pt} 
        \caption{The value of $E_{eq}$ under proposed CSI-free scheme with various $N$ and $\beta_\text{min}$. Regardless of $\beta_{\text{min}}$, $E_{eq}$ is proportional to $N^2$. And when $\beta_\mathrm{min}$ goes to 0.2, 0.5, 0.8, 1.0, take 0.3675, 0.534, 0.7851 and 1.0 for $\tau$, respectively.}
        \label{fig:exam2}
    \end{figure}

    Fig. \ref{fig:exam2} shows that the value of $E_{eq}$ increases with the growth of the number of IRS elements $N$. To better quantify the performance loss caused by imperfect hardware, we fit the curve with $f_{\text{uv}}(N)=E_{eq}=\tau N^2$. As depicted in Fig. \ref{fig:exam2}, the value of $E_{eq}$ increases as $N$ grows and is proportional to $N^2$, implying that the average harvested energy is also proportional to $N^2$. We obtain the boundary of $\kappa_B$ by substituting $E_{eq}$ in Eq. \eqref{eqn:kboundary} with $f_{\text{uv}}(N)=\tau N^2$, resulting in $\kappa_B=\frac{N-R_\Sigma^\S}{\tau N^2-N}$, and Eq. \eqref{eqn:neq1} can be further expressed as 
    \begin{align}
        \kappa &
        \begin{cases}
            \leq \frac{N-R_\Sigma^\S}{\tau N^2-N}, \text{\emph{DM} outperforms \emph{OM}},\\
            >\frac{N-R_\Sigma^\S}{\tau N^2-N}, \text{\emph{OM} outperforms \emph{DM}}.
        \end{cases}
    \end{align}
    \textcolor{red}{Note that the parameter $(1-\tau)$ represents the proportion of performance loss caused by amplitude and phase coupling compared to the ideal IRS model, while the coupled model and the ideal IRS model are equivalent when $\tau=1$.}
    \begin{Rem}
        For sufficiently large $N$ (i.e., $N=100$), we have $\kappa_B=\frac{N-R_\Sigma^\S}{\tau N^2-N}\approx 0, \forall \tau$. Thus, we can conclude that the \emph{DM} scheme is selected only when the channel experiences Rayleigh fading. Otherwise, the \emph{OM} scheme is preferred. In practical WET scenarios, very few cases exhibit $\kappa=0$; therefore, we will only consider the \emph{OM} scheme in the following section.
    \label{remark:4}
    \end{Rem}

    \subsection{Energy Harvesting Performance}
\begin{figure}
        \centering
        \includegraphics[width=0.9\linewidth]{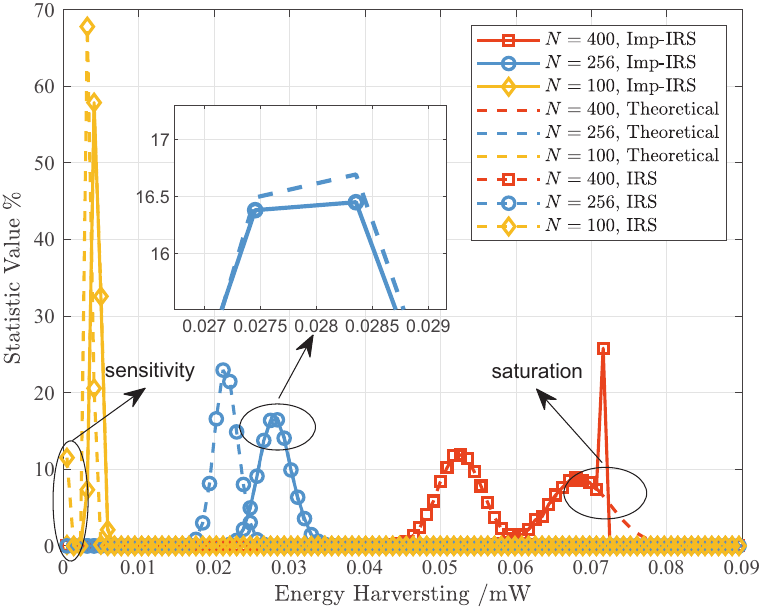}
        \setlength{\abovecaptionskip}{0pt}
        \setlength{\belowcaptionskip}{0pt} 
        \caption{Energy harvesting performance at 4 meters away from IRS with determined direction $\hat{\vartheta}_{h,v}=0$ and Rician factor $\kappa=2$. The number of IRS elements is 100, 256 and 400 respectively.}
        \label{fig:exam3}
    \end{figure}

    In this subsection, we evaluate the energy harvesting performance. We initially set the beam to point in a fixed direction $\hat{\vartheta}_{h,v}=0$ with Rician factor $\kappa=2$, and discuss the energy performance in that direction without applying rotation. Fig. \ref{fig:exam3} presents the distribution of received energy at a distance of 4 meters from the IRS, based on Monte Carlo simulations. The \emph{IRS} case represents a scenario where only phase adjustment is considered, without fully accounting for attenuation due to imperfect hardware during the optimization of the Practical IRS. In contrast, \emph{Imp-IRS} carefully take into account the influence of phase adjustments on the amplitude. Additionally, we take the distribution of received energy shown in Eq. \eqref{eqn:noncentralChi-sqrt} as the theoretical value. We observe that the mean value increases as $N$ grows, as indicated by Eq. \eqref{eqn:expressionMean}. Moreover, the variance derived as
    \begin{align}
        \mathbb{D}\left[{E_k}\right] = \left(\frac{P_eR^\S_{\Sigma}}{2(\kappa+1)}\right)^2\left(4+\frac{8\kappa E_{eq}}{R^\S_{\Sigma}}\right),
    \end{align}
    also increases with $N$.

\begin{figure}
        \centering
        \includegraphics[width=0.9\linewidth]{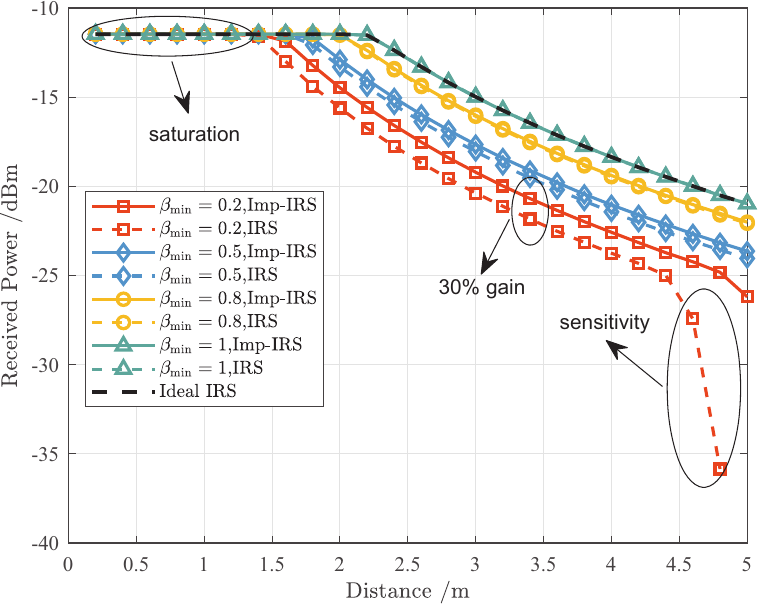}
        \setlength{\abovecaptionskip}{0pt}
        \setlength{\belowcaptionskip}{0pt} 
        \caption{Energy harvesting performance in relation to distance from IRS to ER and various coupling parameters with determined direction $\hat{\vartheta}_{h,v}=0$ and Rician factor $\kappa=2$.}
        \label{fig:exam3_2}
    \end{figure}

    Fig. \ref{fig:exam3_2} illustrates the energy harvesting trend as a function of distance for various coupling parameters, with a fixed direction $\hat{\vartheta}_{h,v}=0$ and Rician factor $\kappa=2$. Note that neglecting imperfect hardware leads to a 30\% decline in performance. This decline increases with the number of IRS elements, as shown in Fig. \ref{fig:exam3}, but decreases as the hardware improves until it is eliminated, as illustrated in Fig. \ref{fig:exam3_2}. Simultaneously, we observe that when a portion of the harvested energy distribution falls below the sensitivity, the average energy decreases rapidly, and the average energy will not reach zero until the entire distribution is below the sensitive.
    \subsection{CSI-free Versus CSI-based}
    \textcolor{red}{In this subsection, we integrate the rotation scheme to further compare the performance of the CSI-based and the proposed CSI-free schemes. It is important to highlight that the CSI-based counterpart operates under the assumption of perfect CSI, and we consider the Ideal IRS model to simplify the CSI-based scheme. As discussed in Remark \ref{remark:1} and Remark \ref{remark:3}, we can still employ MRT as the optimal precoder for the CSI-based counterpart, and the incident phase introduced by MRT can be ignored, as mentioned at the end of Section \ref{sec:IdeaIRS}. The max-min problem of the CSI-based scheme can then be formulated as a semi-definite program (SDP), resulting in
    \begin{align}
        \max_{\mathbf{V}}\quad &t\label{eqn:eqnPcsibased}\\
        \text{s.t.} \quad &\text{trace}\left(\mathbf{V}\mathbf{H}_k\right)\geq t,\forall k\in S,\tag{\ref{eqn:eqnPcsibased}a}\\
        & \text{diag}(\mathbf{V})=\mathbf{1}_N,\tag{\ref{eqn:eqnPcsibased}b}\\
        & \mathbf{V}\succeq\mathbf{0},\tag{\ref{eqn:eqnPcsibased}c}
    \end{align}
    where $\mathbf{V}=\text{diag}(\mathbf{\Theta})\text{diag}(\mathbf{\Theta})^\mathrm{H}\in\mathbb{C}^{N\times N}$, $\mathbf{H}_k=\mathbf{h}_k\mathbf{h}^\mathrm{H}_k\in\mathbb{C}^{N\times N}$, $t$ is an auxiliary variable, and $S$ denotes the set of ERs. Owing to the elimination of the rank-one constraint of $\mathbf{V}$, additional steps (i.e., Gaussian Randomization) are needed to construct a rank-one solution from a higher-rank solution.}
    
\begin{figure}
        \centering

        \subfigure[Beam pattern]{  % []内可单独为每个小图命名。默认按照(a)(b)...的顺序命名，若省去[]则小图不命名。
            \includegraphics[width=0.65\linewidth]{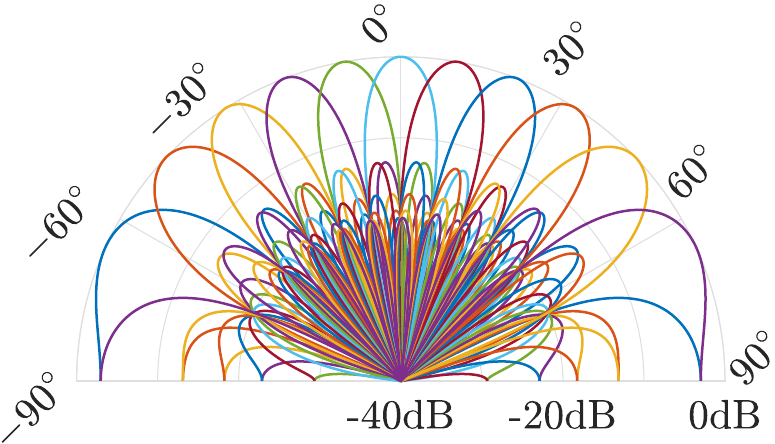}\label{fig:receivedenergy}}
            \setlength{\abovecaptionskip}{0pt}
            \setlength{\belowcaptionskip}{0pt} 
        \subfigure[Entire energy coverage]{
            \includegraphics[width=0.65\linewidth]{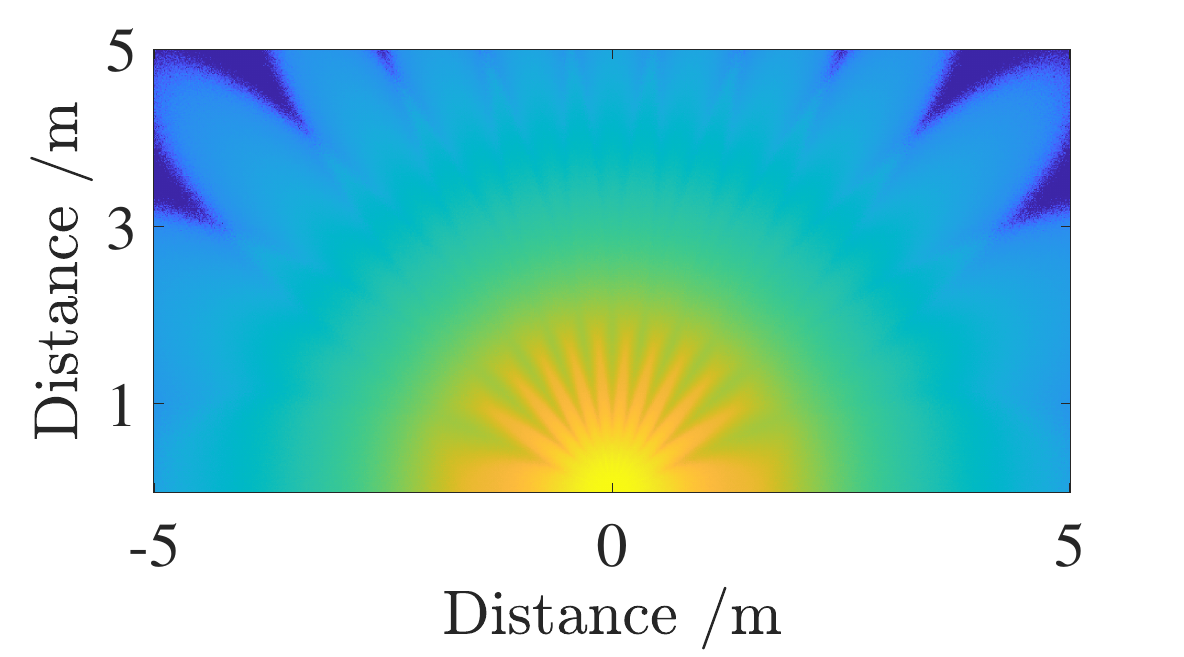}\label{fig:speceff}}
            \setlength{\abovecaptionskip}{0pt}
            \setlength{\belowcaptionskip}{0pt} 
        \caption{Beam pattern and entire energy coverage with 11 directions with $\kappa=2$.}
        \label{fig:HeatmapandBeampattern}
    \end{figure}

    As a counterpart, we choose to rotate in $V=11$ directions within a period using our rotation scheme, as discussed in Sections \ref{sec:beamrotary_perfect} and \ref{sec:beamrotary_imperfect}. The beam directions $\hat{\vartheta}_{h,v}$, where $v \in V$, are as follows: [$-65.7^\circ$, $-46.5^\circ$, $-32.7^\circ$, $-21.0^\circ$, $-10.3^\circ$, $0^\circ$, $10.3^\circ$, $21.0^\circ$, $32.7^\circ$, $46.5^\circ$, $65.7^\circ$]. \textcolor{red}{Note that compared with our proposed CSI-free scheme, the CSI-based counterpart only obtains one optimal $\mathbf{\Theta}$ in a charging cycle and transmits power simultaneously to multiple ERs within the coverage area.} The beam pattern and the entire energy coverage $\mathbf{H}_{cov}$, as described in Eq. \eqref{eqn:heatmapExpression}, are depicted in Fig. \ref{fig:HeatmapandBeampattern}.

    As a performance evaluation criterion, we select the ER with the worst energy harvesting performance from the set $S$ (hereinafter referred to as the worst case) to assess the performance of the CSI-based and CSI-free schemes.

    We know that as the number of IRS elements increases, pilot overhead consumes a substantial portion of the channel correlation time, leading to a significant reduction in available energy transmission time \cite{pilotoverhead1, pilotoverhead2}. To model this phenomenon, we adopt a simple linear model where the pilot overhead $T_p$ increases proportionally with the number of IRS elements (i.e., $T_p = N + 1$) \cite{pilot_linear}. Specifically, we assume that $\mathbf{h}_k$ and $\mathbf{G}$ remain constant over a channel coherence block of length $T_c = 196$, and $T_p < T_c$. Consequently, the remaining $(T_c - T_p = T_c - N - 1)$ is utilized for energy transfer.

\begin{figure}
        \centering
        \includegraphics[width=0.9\linewidth]{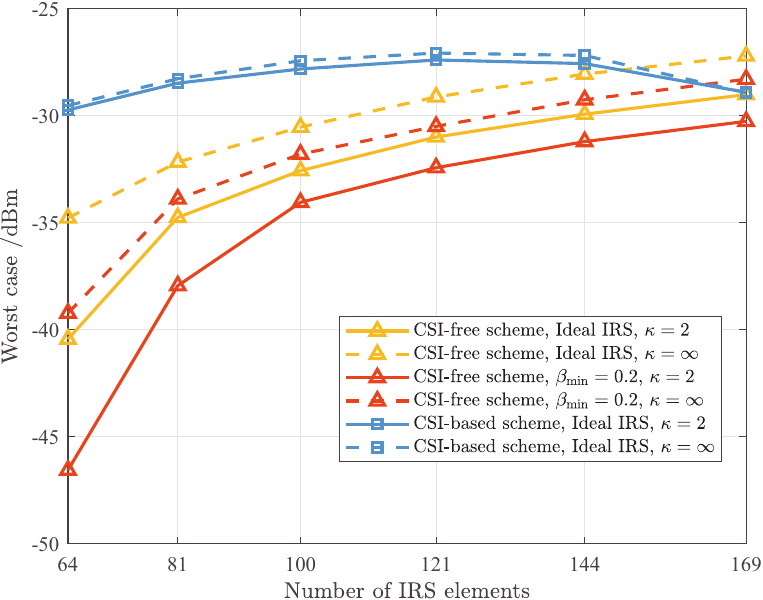}
        \setlength{\abovecaptionskip}{0pt}
        \setlength{\belowcaptionskip}{0pt} 
        \caption{Worst case performance of CSI-based scheme and CSI-free scheme with various $N$.}
        \label{fig:exam4_2}
    \end{figure}

    Fig. \ref{fig:exam4_2} illustrates the performance with $|S| = 16$ ERs and various $N$ values. As the number of IRS elements increases, the CSI-based scheme initially provides more energy to ERs. However, as pilot overhead consumes excessive coherent time, the CSI-based scheme's performance gradually declines. These results also indicate that there exists an optimal number of IRS elements to balance pilot overhead and passive beam gain for CSI-based IRS-assisted WET.

    Based on the aforementioned findings, the CSI-based scheme constrains the increase in the number of IRS elements, thereby limiting IRS-assisted WET performance. In contrast, the performance of the proposed CSI-free scheme increases as $N$ grows, as expected. Moreover, when the number of IRS elements is sufficiently large, our proposed CSI-free scheme outperforms the CSI-based scheme, suggesting that our CSI-free approach is better suited for large-scale IRS with massive elements in future deployments.

\begin{figure}
        \centering
        \includegraphics[width=0.9\linewidth]{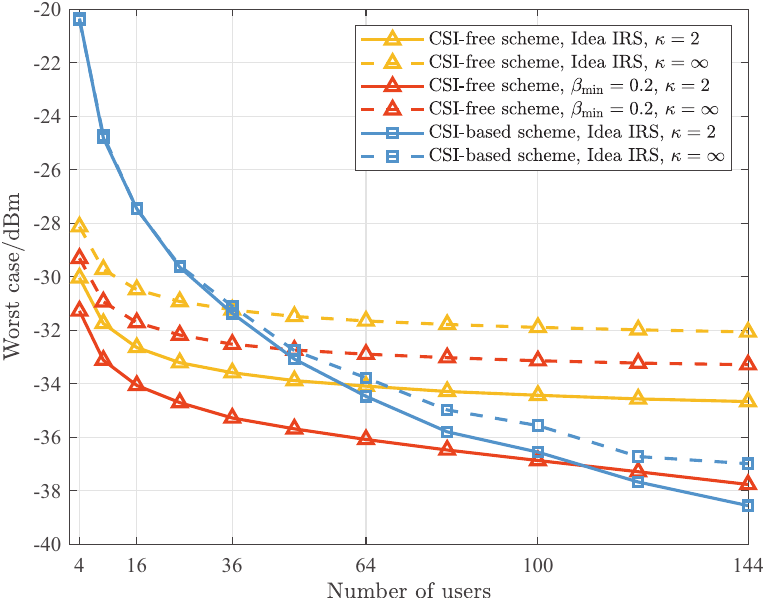}
        \setlength{\abovecaptionskip}{0pt}
        \setlength{\belowcaptionskip}{0pt} 
        \caption{Worst case performance of CSI-based scheme and CSI-free scheme with various number of ERs $|S|$.}
        \label{fig:exam4_1}
    \end{figure}

    Taking pilot overhead into account, we further discuss the impact of the number of ERs. Fig. \ref{fig:exam4_1} demonstrates the performance differences between the CSI-free and CSI-based schemes. Due to the additional channel information, the CSI-based scheme performs considerably better than our CSI-free scheme when the number of ERs is small. \textcolor{red}{However, as the number of ERs increases, the CSI-based scheme fails to accommodate all ERs, leading to a substantial performance decline. Numerical results indicate that our CSI-free solution is highly advantageous for scenarios with a massive number of ERs, especially in future 6G environments where the number of IoT devices per square meter is expected to reach tens or more. In applications such as smart factories and smart homes, where numerous ERs require regular charging and energy demand is not sudden or instantaneous, our CSI-free scheme excels. Nonetheless, the CSI-based scheme remains superior for scenarios with a small number of ERs requiring immediate and sudden charging.} To demonstrate the impact of imperfect hardware, we also compare the performance differences between Practical and Ideal IRS under the CSI-free scheme. However, it is important to emphasize that such comparisons are not entirely fair, as they involve different hardware.

    \section{Conclusion}\label{sec:5}
    In this paper, we proposed a CSI-free scheme capable of supporting large-scale IRS elements and massive ERs without compromising energy harvesting performance, without altering the existing IRS hardware architecture. Initially, we developed our CSI-free scheme for the uncoupled reflection amplitude and phase shift IRS model and subsequently extended it to the coupled reflection amplitude and phase shift IRS model. Employing a phased beam rotation scheme, our approach achieves full spatial energy coverage within a single rotation. Furthermore, extensive simulations were conducted to demonstrate the superiority of our CSI-free scheme, particularly in scenarios involving massive ERs or large IRS elements, where our CSI-free approach significantly outperforms the CSI-based scheme. In essence, our scheme allows for an arbitrary increase in the number of IRS elements without negatively impacting performance due to excessive pilot overhead. Simultaneously, our method incurs no cost for adding or removing ERs, rendering it more suitable for practical applications in the upcoming IoT era.
    {\appendices
    \section{Proof of Lemma 1}\label{app:A}
    As we mentioned early, the PB-to-IRS channel $\mathbf{G}$ can be expressed as
    \begin{align}
        \mathbf{G}=&\sqrt{MN}\boldsymbol{\alpha}_{G,r}(\vartheta_{G},\varphi_{G})\boldsymbol{\alpha}_{G,t}(\gamma_G)^\mathrm{H} \nonumber\\
        =&\sqrt{MN}\left[
        [\boldsymbol{\alpha}_{G,x}(u_G)]_1\boldsymbol{\alpha}_{G,y}(v_G)\boldsymbol{\alpha}_{G,t}(z_G)^\mathrm{H},
        \cdots,\right.\nonumber\\
        &\left.[\boldsymbol{\alpha}_{G,x}(u_G)]_{N_x-1}\boldsymbol{\alpha}_{G,y}(v_G)\boldsymbol{\alpha}_{G,t}(z_G)^\mathrm{H}
        \right]^\mathrm{T}.
    \end{align}
    Therefore, the $j$-th row of $\mathbf{G}$ can be expressed as
    \begin{align}
        \sqrt{MN}[\boldsymbol{{\alpha}}_{G,y}(v_G)]_{\text{mod}(j,N_y)}[\boldsymbol{{\alpha}}_{G,x}(u_G)]_{(\lfloor j/N_x\rfloor+1)}\boldsymbol{\alpha}_{G,t}(z_G)^\mathrm{H}.\label{eqn:channelG}
    \end{align}
    Thus, the $j$-th element's incident signal is obtained as
    \begin{align}
        \mathcal{A}_j=&\mathbf{G}_{j,:}\sum\limits_{i=1}^N\mathbf{w}_i\nonumber\\
        =&\sqrt{MN}[\boldsymbol{{\alpha}}_{G,y}(v_G)]_{\text{mod}(j,N_y)}[\boldsymbol{{\alpha}}_{G,x}(u_G)]_{(\lfloor j/N_x\rfloor+1)}\nonumber\\
        &\times\boldsymbol{\alpha}_{G,t}(z_G)^\mathrm{H}\sum\limits_{i=1}^N\mathbf{w}_i.
    \end{align}

    Due to the fact that the first two components associated with index $j$ are only related to phase and are independent of amplitude, whereas the last two components are common; thus, the incident signal of each element has the same amplitude but different phase.
    \section{Proof of Lemma 2}\label{app:B}
    Assume that there are two random variables $\mathcal{X}\sim \mathcal{U}(-\pi,\pi)$ and $\mathcal{Y}\sim \mathcal{U}(-\pi,\pi)$. The Cumulative Distribution Function (CDF) of random variable $\mathcal{Z}=\mathcal{X}-\mathcal{Y}$ can be expressed as
    \begin{align}
        \mathbb{P}(\mathcal{Z}\leq z)=\mathbb{P}(\mathcal{X}-\mathcal{Y}\leq z), \,\,z\in[-2\pi,2\pi],
    \end{align}
    and while $z\in[-2\pi,0)$, we have
    \begin{align}
        \iint\limits_{\mathbb{D}}f(x,y)dxdy&=\int_{-\pi}^{\pi+z}dx\int_{x-z}^\pi f(x,y)dy\nonumber\\
        & = \frac{1}{8\pi^2}z^2+\frac{1}{2\pi}z+\frac{1}{2},
    \end{align}
    where $f(x,y)=f(x)f(y)=1/4\pi^2$ since $\mathcal{X}$ and $\mathcal{Y}$ are independent.
    Similarly, while $z\in[0,2\pi]$, we have
    \begin{align}
        &\iint\limits_{\mathbb{D}_1}f(x,y)dxdy+\iint\limits_{\mathbb{D}_2}f(x,y)dxdy\nonumber\\
        &=\int_{-\pi}^{z-\pi}dx\int_{-\pi}^\pi f(x,y)dy+\int_{z-\pi}^{\pi}dx\int_{x-z}^\pi f(x,y)dy\nonumber\\
        &=-\frac{1}{8\pi^2}z^2+\frac{1}{2\pi}z+\frac{1}{2}.
    \end{align}
    Thus we have the Probability Density Function (PDF) of $\mathcal{Z}$ as
    \begin{align}
        f(z) &=
        \begin{cases}
            -\frac{1}{4\pi^2}z+\frac{1}{2\pi}, \,\, z\in[0, 2\pi],\\
            \frac{1}{4\pi^2}z+\frac{1}{2\pi}, \,\, z\in[-2\pi, 0).
        \end{cases}
        \label{eqn:triangDistrubution}
    \end{align}
    Eq. \eqref{eqn:triangDistrubution} shows that the subtraction of two random variables with uniform distribution yields triangular distribution. Furthermore, since the phase should be limited to $[-\pi,\pi]$, we have
    \begin{align}
        f(z) &\overset{(a)}=
        \begin{cases}
            -\frac{1}{4\pi^2}z+\frac{1}{2\pi}+(\frac{1}{4\pi^2}(z-2\pi)+\frac{1}{2\pi}), \,\, z\in[0, \pi],\nonumber\\
            \frac{1}{4\pi^2}z+\frac{1}{2\pi}+(-\frac{1}{4\pi^2}(z+2\pi)+\frac{1}{2\pi}), \,\, z\in[-\pi, 0),
        \end{cases}\nonumber\\
        &= \frac{1}{2\pi}, \,\, z\in[-\pi, \pi],
        \label{eqn:triang2Uniform}
    \end{align}
    where $(a)$ is the outcome of shifting the PDF in $[-2\pi,-\pi]$ and $[\pi, 2\pi]$ to $[-\pi, \pi]$. Eq. \eqref{eqn:triang2Uniform} illustrates that the subtraction of two phases that obey the uniform distribution from $-\pi$ to $\pi$ is still the uniform distribution from $-\pi$ to $\pi$.

    Due to $\Phi_{k,t},\forall t\in N$ ($\mu_t,\forall t\in N$) change in the index of the IRS element (i.e., $\Phi_{k,t}=-(\text{mod}(t,N_y)-1)\pi\sin\vartheta_{h,k},\forall t\in N$), it can be regarded as a uniform random variable. Thus we have both $(\Phi_{k,t}-\Phi_{k,l})$ and ($\mu_l-\mu_t$) obey uniform distribution according to the preceding conclusion. Then using the conclusion again, we finally have the results that $(\Phi_{k,t}-\Phi_{k,l} - (\mu_l-\mu_t) = \Phi_{k,t}+\mu_t-\Phi_{k,l}-\mu_l)$ also follows uniform distribution. Obviously, we have $2\sum_{t=1}^{N-1}\sum_{l=t+1}^N\cos(\Phi_{k,t}+\mu_t-\Phi_{k,l}-\mu_l)\approx 0$. Thus the proof is completed.

    \section{Proof of Lemma 3}\label{app:C}
    In order to prove that MRT in Remark \ref{remark:1} is still the optimal solution of problem $\text{(P2)}$, we first analyze how MRT maximizes the incident power of IRS elements. We assume MRT can be expressed as $\mathbf{w}_{\mathrm{MRT}}=\left[
        \begin{array}{c c c c}
        \mathcal{C}_1e^{\mathbbm{i}\zeta_1}& \mathcal{C}_2e^{\mathbbm{i}\zeta_2}&\cdots& \mathcal{C}_Me^{\mathbbm{i}\zeta_M}
        \end{array}
        \right]^\mathrm{H}$
    where $\mathcal{C}_j,\forall j\in M$ and $\zeta_j, \forall j\in M$ denote the transmit amplitude and phase shift on $j$-th antenna, respectively. According to Lemma 1 and Remark 1, maximizing the incident signal power on the element of IRS is equivalent to maximizing
    \begin{align}
        &\sqrt{M}\boldsymbol{\alpha}_{G,t}(z_G)^\mathrm{H}\mathbf{w}_\mathrm{MRT}\nonumber\\
        =&\left[
        \begin{array}{ccc}
        1&\cdots&e^{\mathbbm{i}(M-1)z_G}
        \end{array}
        \right]^\mathrm{\dagger}
        \left[
        \begin{array}{ccc}
        \mathcal{C}_1e^{\mathbbm{i}\zeta_1}&\cdots&\mathcal{C}_Me^{\mathbbm{i}\zeta_M}
        \end{array}
        \right]^\mathbf{H}\nonumber\\
        =&\mathcal{C}_1e^{-\mathbbm{i}\zeta_1}+\mathcal{C}_2e^{-\mathbbm{i}(\zeta_2+z_G)}+\cdots+\mathcal{C}_Me^{-\mathbbm{i}(\zeta_M+(M-1)z_G)},\label{eqn:proo2aW}
    \end{align}
    and MRT enables phase alignment, with $\zeta_1=\cdots=\zeta_M+(M-1)z_G$. Moreover, the transmit power allocation of each antenna is obtained easily by Cauchy-Schwarz inequality as $\mathcal{C}_1=\mathcal{C}_2=,\cdots,=\mathcal{C}_M=\sqrt{P/M}$ under transmit power constraint $\sum_{m=1}^M\mathcal{C}_m^2=P$.

    Assume there is a precoder $\mathbf{w}^*$ that outperforms $\mathbf{w}_{\mathrm{MRT}}$, which can be expressed as
    \begin{align}
        \mathbf{w}^*=\text{diag}(\boldsymbol{\xi})\mathbf{w}_\mathrm{MRT},
    \end{align}
    where $\boldsymbol{\xi}=[e^{-\mathbbm{i}\xi_1},\cdots,e^{-\mathbbm{i}\xi_M}]$, and Eq. \eqref{eqn:proo2aW} can be rewritten as
    \begin{align}
    \sqrt{M}\boldsymbol{{\alpha}}_{G,t}(z_G)^\mathrm{H}\mathbf{w}^*&=\mathcal{C}_1e^{-\mathbbm{i}(\zeta_1+\xi_1)}+\cdots+\mathcal{C}_Me^{-\mathbbm{i}(\zeta_M+(M-1){z_G}+\xi_M)}\nonumber\\
    &=\sqrt{M}\boldsymbol{{\alpha}}_{G,t}(z_G)^\mathrm{H}\mathbf{w}_\mathrm{MRT}\frac{e^{-\mathbbm{i}\xi_1}+\cdots+e^{-\mathbbm{i}\xi_M}}{M}\nonumber\\
    &\overset{(a)}{=}\sqrt{M}\boldsymbol{{\alpha}}_{G,t}(z_G)^\mathrm{H}\mathbf{w}_\mathrm{MRT}\frac{\mathcal{C}_\Sigma}{M}e^{\mathbbm{i}\xi_{\Sigma}},
    \label{eqn:remainordecrease}
    \end{align}
    where $(a)$ comes from let $\sum_{j=1}^Me^{-\mathbbm{i}\xi_j}=\mathcal{C}_\Sigma e^{\mathbbm{i}\xi_{\Sigma}}$, and we can easily observe that $\mathcal{C}_\Sigma/M\leq 1$. It shows the fact that $\mathbf{w}^*$ will decrease (or remain) the incident power on the IRS element. And considering the additional incident phase of $i$-th IRS element according to Remark \ref{remark:1}, we have
    \begin{align}
        &\mu^*_i-\mu_i\nonumber\\
        &\overset{(a)}{=}\arg([\boldsymbol{{\alpha}}_{G,y}(v_G)]_{\text{mod}(i,N_y)}[\boldsymbol{{\alpha}}_{G,x}(u_G)]_{(\lfloor i/N_x\rfloor+1)}\boldsymbol{\alpha}_{G,t}(z_G)^\mathrm{H}\mathbf{w}_\mathrm{MRT})\nonumber\\
        &\quad-\arg([\boldsymbol{{\alpha}}_{G,y}(v_G)]_{\text{mod}(i,N_y)}[\boldsymbol{{\alpha}}_{G,x}(u_G)]_{(\lfloor i/N_x\rfloor+1)}\boldsymbol{\alpha}_{G,t}(z_G)^\mathrm{H}\mathbf{w}^*)\label{eqn:thesamecomp}\\
        &\overset{(b)}{=}\arg(\boldsymbol{\alpha}_{G,t}(z_G)^\mathrm{H}\mathbf{w}_\mathrm{MRT})-\arg(\boldsymbol{\alpha}_{G,t}(z_G)^H\mathbf{w}^*)\nonumber\\
        &\overset{(c)}{=}\xi_{\Sigma},\forall i\in N.\label{eqn:proof2samephase}
    \end{align}
    where $\mu^*$ ($\mu$) denotes the incident phase using $\mathbf{w}_\mathrm{MRT}$ ($\mathbf{w}^*$) as mentioned in Eq. \eqref{eqn:receiveEk}, and $(a)$ comes from Eq. \eqref{eqn:channelG}, $(b)$ follows the fact that the first two components of Eq. \eqref{eqn:thesamecomp} are the same. And (c) follows Eq. \eqref{eqn:remainordecrease}.

    Eq. \eqref{eqn:proof2samephase} illustrates that compared with $\mathbf{w}_{\mathrm{MRT}}$, $\mathbf{w}^*$ will bring the same phase $\xi_{\Sigma}$ to each IRS element.

    According to Eq. \eqref{eqn:ignoreRsigma}, we have $\mu^*_t-\mu^*_l = \mu_t+\xi_\Sigma-(\mu_l+\xi_\Sigma)=\mu_t-\mu_l$ and obviously, the additional phase $\xi_\Sigma$ will not change the value of $E_{eq}$.

    In conclusion, precoder $\mathbf{w}^*$ will decrease (or remain) the incident power of the $i$-th element, and does not affect the maximal value of $E_{eq}$, which illustrates that there is no better precoder than $\mathbf{w}_{\text{MRT}}$. Thus, we can conclude that $\mathbf{w}_\mathrm{MRT}$ is the optimal precoder for $\text{(P2)}$.
    }

	\bibliography{Reference}

\begin{thebibliography}{10}
\providecommand{\url}[1]{#1}
\csname url@samestyle\endcsname
\providecommand{\newblock}{\relax}
\providecommand{\bibinfo}[2]{#2}
\providecommand{\BIBentrySTDinterwordspacing}{\spaceskip=0pt\relax}
\providecommand{\BIBentryALTinterwordstretchfactor}{4}
\providecommand{\BIBentryALTinterwordspacing}{\spaceskip=\fontdimen2\font plus
\BIBentryALTinterwordstretchfactor\fontdimen3\font minus
  \fontdimen4\font\relax}
\providecommand{\BIBforeignlanguage}[2]{{%
\expandafter\ifx\csname l@#1\endcsname\relax
\typeout{** WARNING: IEEEtran.bst: No hyphenation pattern has been}%
\typeout{** loaded for the language `#1'. Using the pattern for}%
\typeout{** the default language instead.}%
\else
\language=\csname l@#1\endcsname
\fi
#2}}
\providecommand{\BIBdecl}{\relax}
\BIBdecl

\bibitem{intro_overviewbruno1}
B.~Clerckx, K.~Huang, L.~R. Varshney, S.~Ulukus, and M.-S. Alouini, ``Wireless
  power transfer for future networks: Signal processing, machine learning,
  computing, and sensing,'' \emph{IEEE Journal of Selected Topics in Signal
  Processing}, vol.~15, no.~5, pp. 1060--1094, 2021.

\bibitem{intro_overviewbruno2}
B.~Clerckx, J.~Kim, K.~W. Choi, and D.~I. Kim, ``Foundations of wireless
  information and power transfer: Theory, prototypes, and experiments,''
  \emph{Proceedings of the IEEE}, vol. 110, no.~1, pp. 8--30, 2022.

\bibitem{SWIPT3}
B.~Clerckx, ``Wireless information and power transfer: Nonlinearity, waveform
  design, and rate-energy tradeoff,'' \emph{IEEE Transactions on Signal
  Processing}, vol.~66, no.~4, pp. 847--862, 2018.

\bibitem{intro_Mon}
F.~A. Monteiro, O.~L.~A. López, and H.~Alves, ``Massive wireless energy
  transfer with statistical {CSI} beamforming,'' \emph{IEEE Journal of Selected
  Topics in Signal Processing}, vol.~15, no.~5, pp. 1169--1184, 2021.

\bibitem{huj6G}
J.~Hu, Q.~Wang, and K.~Yang, ``Energy self-sustainability in full-spectrum
  6{G},'' \emph{IEEE Wireless Communications}, vol.~28, no.~1, pp. 104--111,
  2021.

\bibitem{intro_overviewonel1}
O.~L.~A. López, H.~Alves, R.~D. Souza, S.~Montejo-Sánchez, E.~M.~G.
  Fernández, and M.~Latva-Aho, ``Massive wireless energy transfer: Enabling
  sustainable {IoT} toward 6{G} era,'' \emph{IEEE Internet of Things Journal},
  vol.~8, no.~11, pp. 8816--8835, 2021.

\bibitem{mahmood2020white}
N.~H. Mahmood, S.~B{\"o}cker, A.~Munari, F.~Clazzer, I.~Moerman, K.~Mikhaylov,
  O.~Lopez, O.-S. Park, E.~Mercier, H.~Bartz \emph{et~al.}, ``White paper on
  critical and massive machine type communication towards {6G},'' \emph{arXiv
  preprint arXiv:2004.14146}, 2020.

\bibitem{lopez2020csi}
O.~L. L{\'o}pez, S.~Montejo-S{\'a}nchez, R.~D. Souza, C.~B. Papadias, and
  H.~Alves, ``On {CSI}-free multiantenna schemes for massive {RF} wireless
  energy transfer,'' \emph{IEEE Internet of Things Journal}, vol.~8, no.~1, pp.
  278--296, 2020.

\bibitem{zhangBX1}
B.~Zhang, K.~Wang, K.~Yang, and G.~Zhang, ``{IRS}-assisted short packet
  wireless energy transfer and communications,'' \emph{IEEE Wireless
  Communications Letters}, vol.~11, no.~2, pp. 303--307, 2022.

\bibitem{SWIPT1}
J.~Xu, L.~Liu, and R.~Zhang, ``Multiuser {MISO} beamforming for simultaneous
  wireless information and power transfer,'' \emph{IEEE Transactions on Signal
  Processing}, vol.~62, no.~18, pp. 4798--4810, 2014.

\bibitem{YueSWIPT}
Q.~Yue, J.~Hu, K.~Yang, and C.~Huang, ``Transceiver design for simultaneous
  wireless information and power multicast in multi-user mmwave mimo system,''
  \emph{IEEE Transactions on Vehicular Technology}, vol.~69, no.~10, pp.
  11\,394--11\,407, 2020.

\bibitem{WPCN1}
H.~Ju and R.~Zhang, ``Throughput maximization in wireless powered communication
  networks,'' \emph{IEEE Transactions on Wireless Communications}, vol.~13,
  no.~1, pp. 418--428, 2014.

\bibitem{WPBC1}
C.~Boyer and S.~Roy, ``— invited paper — backscatter communication and
  {RFID}: Coding, energy, and {MIMO} analysis,'' \emph{IEEE Transactions on
  Communications}, vol.~62, no.~3, pp. 770--785, 2014.

\bibitem{IRSassistedSWIPT}
W.~Shi, X.~Zhou, L.~Jia, Y.~Wu, F.~Shu, and J.~Wang, ``Enhanced secure wireless
  information and power transfer via intelligent reflecting surface,''
  \emph{IEEE Communications Letters}, vol.~25, no.~4, pp. 1084--1088, 2021.

\bibitem{IRSassistedSWIPT2}
C.~Pan, H.~Ren, K.~Wang, M.~Elkashlan, A.~Nallanathan, J.~Wang, and L.~Hanzo,
  ``Intelligent reflecting surface aided {MIMO} broadcasting for simultaneous
  wireless information and power transfer,'' \emph{IEEE Journal on Selected
  Areas in Communications}, vol.~38, no.~8, pp. 1719--1734, 2020.

\bibitem{IRSassistedSWIPT5}
Y.~Zhao, B.~Clerckx, and Z.~Feng, ``{IRS}-aided {SWIPT}: Joint waveform, active
  and passive beamforming design under nonlinear harvester model,'' \emph{IEEE
  Transactions on Communications}, vol.~70, no.~2, pp. 1345--1359, 2022.

\bibitem{IRSassistedWPCN}
M.~Hua and Q.~Wu, ``Joint dynamic passive beamforming and resource allocation
  for {IRS}-aided full-duplex {WPCN},'' \emph{IEEE Transactions on Wireless
  Communications}, vol.~21, no.~7, pp. 4829--4843, 2022.

\bibitem{IRS_WET}
Q.~Wu and R.~Zhang, ``Weighted sum power maximization for intelligent
  reflecting surface aided {SWIPT},'' \emph{IEEE Wireless Communications
  Letters}, vol.~9, no.~5, pp. 586--590, 2020.

\bibitem{IRSassistedSWIPT3}
------, ``Joint active and passive beamforming optimization for intelligent
  reflecting surface assisted {SWIPT} under {Q}o{S} constraints,'' \emph{IEEE
  Journal on Selected Areas in Communications}, vol.~38, no.~8, pp. 1735--1748,
  2020.

\bibitem{wqq_survey}
Q.~Wu, S.~Zhang, B.~Zheng, C.~You, and R.~Zhang, ``Intelligent reflecting
  surface-aided wireless communications: A tutorial,'' \emph{IEEE Transactions
  on Communications}, vol.~69, no.~5, pp. 3313--3351, 2021.

\bibitem{pilotoverhead1}
K.~Zhi, C.~Pan, H.~Ren, and K.~Wang, ``Power scaling law analysis and phase
  shift optimization of {RIS}-aided massive {MIMO} systems with statistical
  {CSI},'' \emph{IEEE Transactions on Communications}, vol.~70, no.~5, pp.
  3558--3574, 2022.

\bibitem{pilotoverhead2}
C.~Pan, G.~Zhou, K.~Zhi, S.~Hong, T.~Wu, Y.~Pan, H.~Ren, M.~D. Renzo,
  A.~Lee~Swindlehurst, R.~Zhang, and A.~Y. Zhang, ``An overview of signal
  processing techniques for {RIS/IRS}-aided wireless systems,'' \emph{IEEE
  Journal of Selected Topics in Signal Processing}, vol.~16, no.~5, pp.
  883--917, 2022.

\bibitem{taha2021enabling}
A.~Taha, M.~Alrabeiah, and A.~Alkhateeb, ``Enabling large intelligent surfaces
  with compressive sensing and deep learning,'' \emph{IEEE access}, vol.~9, pp.
  44\,304--44\,321, 2021.

\bibitem{zhu2022sensing}
J.~Zhu, K.~Liu, Z.~Wan, L.~Dai, T.~J. Cui, and H.~V. Poor, ``Sensing riss:
  Enabling dimension-independent csi acquisition for beamforming,'' \emph{IEEE
  Transactions on Information Theory}, vol.~69, no.~6, pp. 3795--3813, 2023.

\bibitem{ren2021configuring}
S.~Ren, K.~Shen, Y.~Zhang, X.~Li, X.~Chen, and Z.-Q. Luo, ``Configuring
  intelligent reflecting surface with performance guarantees: Blind
  beamforming,'' \emph{IEEE Transactions on Wireless Communications}, vol.~22,
  no.~5, pp. 3355--3370, 2023.

\bibitem{QintaoCSIfree}
Q.~Tao, S.~Zhang, C.~Zhong, and R.~Zhang, ``Intelligent reflecting surface
  aided multicasting with random passive beamforming,'' \emph{IEEE Wireless
  Communications Letters}, vol.~10, no.~1, pp. 92--96, 2021.

\bibitem{locinfo}
X.~Hu, C.~Zhong, Y.~Zhang, X.~Chen, and Z.~Zhang, ``Location information aided
  multiple intelligent reflecting surface systems,'' \emph{IEEE Transactions on
  Communications}, vol.~68, no.~12, pp. 7948--7962, 2020.

\bibitem{angleinfo}
\BIBentryALTinterwordspacing
C.~Luo, J.~Hu, L.~Xiang, and K.~Yang, ``Reconfigurable intelligent sensing
  surface aided wireless powered communication networks: A
  sensing-then-reflecting approach,'' 2023. [Online]. Available:
  \url{https://arxiv.org/abs/2310.13335}
\BIBentrySTDinterwordspacing

\bibitem{Rzhang_practicalModel}
S.~Abeywickrama, R.~Zhang, Q.~Wu, and C.~Yuen, ``Intelligent reflecting
  surface: Practical phase shift model and beamforming optimization,''
  \emph{IEEE Transactions on Communications}, vol.~68, no.~9, pp. 5849--5863,
  2020.

\bibitem{ignoreNoisepower1}
Z.~Hou, H.~Chen, Y.~Li, and B.~Vucetic, ``Incentive mechanism design for
  wireless energy harvesting-based internet of things,'' \emph{IEEE Internet of
  Things Journal}, vol.~5, no.~4, pp. 2620--2632, 2018.

\bibitem{ignoreNoisepower2}
S.~Bi, Y.~Zeng, and R.~Zhang, ``Wireless powered communication networks: An
  overview,'' \emph{IEEE Wireless Communications}, vol.~23, no.~2, pp. 10--18,
  2016.

\bibitem{zeng2014optimized}
Y.~Zeng and R.~Zhang, ``Optimized training design for wireless energy
  transfer,'' \emph{IEEE Transactions on Communications}, vol.~63, no.~2, pp.
  536--550, 2014.

\bibitem{tse2005fundamentalsMRT}
D.~Tse and P.~Viswanath, \emph{Fundamentals of wireless communication}.\hskip
  1em plus 0.5em minus 0.4em\relax Cambridge university press, 2005.

\bibitem{wu2019intelligent}
Q.~Wu and R.~Zhang, ``Intelligent reflecting surface enhanced wireless network
  via joint active and passive beamforming,'' \emph{IEEE Transactions on
  Wireless Communications}, vol.~18, no.~11, pp. 5394--5409, 2019.

\bibitem{Beamwidth}
C.~A. Balanis, \emph{Antenna theory: Analysis and design}.\hskip 1em plus 0.5em
  minus 0.4em\relax John wiley \& sons, 2015.

\bibitem{OneLRotaryAntenna}
O.~L.~A. López, H.~Alves, S.~Montejo-Sánchez, R.~D. Souza, and M.~Latva-aho,
  ``{CSI}-free rotary antenna beamforming for massive {RF} wireless energy
  transfer,'' \emph{IEEE Internet of Things Journal}, vol.~9, no.~10, pp.
  7375--7387, 2022.

\bibitem{ke2022power}
X.~Ke, X.~Luping, H.~Jie, and Y.~Kun, ``Power allocation satisfying user
  fairness for integrated sensing and communication system based on orthogonal
  time frequency space modulation.'' \emph{Telecommunications Science},
  vol.~38, no.~9, 2022.

\bibitem{Xiang2023RobustNO}
\BIBentryALTinterwordspacing
L.~Xiang, K.~Xu, J.~Hu, C.~Masouros, and K.~Yang, ``Robust {NOMA}-assisted
  {OTFS}-{ISAC} network design with 3{D} motion prediction topology,'' 2023.
  [Online]. Available: \url{https://arxiv.org/abs/2310.13984}
\BIBentrySTDinterwordspacing

\bibitem{lopez2019statistical}
O.~L. L{\'o}pez, H.~Alves, R.~D. Souza, and S.~Montejo-S{\'a}nchez,
  ``Statistical analysis of multiple antenna strategies for wireless energy
  transfer,'' \emph{IEEE Transactions on Communications}, vol.~67, no.~10, pp.
  7245--7262, 2019.

\bibitem{pilot_linear}
N.~K. Kundu and M.~R. Mckay, ``Large intelligent surfaces with channel
  estimation overhead: Achievable rate and optimal configuration,'' \emph{IEEE
  Wireless Communications Letters}, vol.~10, no.~5, pp. 986--990, 2021.

\end{thebibliography}
\end{document}